\documentclass[pre,superscriptaddress,twocolumn]{revtex4-1}
\pdfoutput=1
\usepackage[colorlinks=true,urlcolor=blue]{hyperref}
\usepackage{amsfonts}
\usepackage{graphicx}
\usepackage{placeins}
\usepackage{amsmath}
\usepackage{xcolor}
\usepackage{bm}
\usepackage{url}

\usepackage{color}
\definecolor{red}{rgb}{0.75,0,0}
\definecolor{blue}{rgb}{0,0,0.75}
\definecolor{green}{rgb}{0,0.5,0}

\begin{document}

\title{Linear response to leadership, effective temperature and decision making in flocks}

\author{D. J. G. Pearce}
\affiliation{Instituut-Lorentz, Universiteit Leiden, P.O. Box 9506, 2300 RA Leiden, The Netherlands}
\author{L. Giomi}
\affiliation{Instituut-Lorentz, Universiteit Leiden, P.O. Box 9506, 2300 RA Leiden, The Netherlands}

\begin{abstract}
Large collections of autonomously moving agents, such as animals or micro-organisms, are able to ``flock'' coherently in space even in the absence of a central control mechanism. While the direction of the flock resulting from this critical behavior is random, this can be controlled by a small subset of informed individuals acting as leaders of the group. In this article we use the Vicsek model to investigate how flocks respond to leadership and make decisions. Using a combination of numerical simulations and continuous modeling we demonstrate that flocks display a linear response to leadership that can be cast in the framework of the fluctuation-dissipation theorem, identifying an ``effective temperature'' reflecting how promptly the flock reacts to the initiative of the leaders. The linear response to leadership also holds in the presence of two groups of informed individuals with competing interests, indicating that the flock's behavioral decision is determined by both the number of leaders and their degree of influence. 
\end{abstract}

\maketitle

\section{Introduction}

The term ``flocking'' (or equivalently swarming, schooling, herding etc.) describes the ability of groups of living organisms to move coherently in space and time \cite{reynolds1987flocks,vicsek2012collective,giardinarev}. This behavior is ubiquitous in nature: it occurs in sub-cellular systems \cite{schaller2010polar}, bacteria \cite{peruani2012collective}, insects \cite{guttal2012cannibalism,bazazi2008collective}, fish \cite{herbert2011inferring,misund1995mapping}, birds \cite{ballerinitopol,pearce2014role,Cavagnaboundaryinfo,bialek2012statistical,pearce2014density} and in general in nearly any group of individuals endowed with the ability to move and sense. This spectacular example of robustness has inspired science and technology in a two-fold way: on the one hand scientists have focused their efforts in understanding the origin of a collective behavior found in systems of such an astonishing diversity \cite{vicsek2012collective}; on the other hand technologists have envisioned the possibility of implementing this form of social organization that spontaneously arises in living systems to construct flocks of devices that can work independently and yet collectively towards a common goal \cite{giomi2013swarming,rubenstein2014programmable}. 

A particularly interesting question in the context of collective behavior in biological and bio-inspired systems revolves around how groups respond to the leadership of a subset of individuals having pertinent information. In animals, such information might represent the location of a food source \cite{couzin2011uninformed}, a specific migration route \cite{dell2008flock}, or a threat of which part of the group is unaware, such as a predator only visible to a minority of individuals \cite{ward2011fast}. In biomimetic systems, on the other hand, this might consist of a set of instructions related to the group task. The response of schooling fish to leadership has represented, in particular, the focus of several empirical studies. This is due to the possibility of training fish to swim toward a specific target, expect food at a given time or location \cite{reebs2000can,krause2000leadership,leblond2006individual} or the ability to insert remote-controlled replica animals \cite{faria2010robofish,ward2008quorum}, thus acting as leaders for the remaining fish. While varying in the details, these studies have demonstrated that large groups of individuals are able to adopt the behavior of an informed subset. The statistical mechanics of leadership and decision making in animal groups has been systematically investigated by Couzin and coworkers in a series of seminal works \cite{couzin2005effective,ward2008quorum,sumpter2008consensus,couzin2011uninformed}. Using a combination of experiments and numerical simulations based on self-propelled particles models, they showed that communities of collectively moving individuals are able to make consensus decisions in the presence of a small minority of unorganized informed individuals. Furthermore, they demonstrated that when two informed subsets with competing behaviors are introduced, the group selects the behavior of the larger informed subset with an accuracy that increases with the number of uninformed individuals \cite{couzin2011uninformed}.

The generality and the robustness of these results have acted as a stimulus to identify a generic mechanism behind leadership and decision making in systems of collectively moving individuals \cite{kao2014collective,guttal2011leadership}. Yet, whether it is possible to identify the basic laws governing the response of a group to leadership, is still unclear. 

In order to gain insight into this problem, we present here a linear response analysis of a model flock whose dynamics is described by the Vicsek model with angular noise \cite{vicsek}. We study how a collection of flocking agents responds to the leadership of a randomly selected subset of the entire flock that is biased to turn toward a specific direction. Using numerical simulations and continuous modeling, we demonstrate that the system's response to leadership can be cast in the framework of the fluctuation-dissipation theorem, upon introducing an ``effective temperature'' proportional to the ratio between the correlation and response functions and generally dependent on the system density, velocity and noise. Remarkably, both the density and velocity dependence disappear at large densities, revealing a universal linear dependence of the effective temperature on the noise variance. We then apply this approach to the case wherein the flock must choose between two subsets of leaders with competing interests, identifying again a linear response to the total perturbation applied by both groups. In this case, however, the flock behavioral decision is determined by both the number of leaders and their degree of influence, so a small subgroup of particularly influential informed individuals can overrule a larger subset of less influential informed individuals.

\section{Results}
\subsection{\label{sec:discrete_flocks}Discrete flocks}

Let us consider the Vicsek model subject to angular white noise \cite{vicsek,czirok1997spontaneously,chate2008collective,ginelli2015physics}. The system consists of $N$ individuals traveling at velocity $\bm{v}_{i}=v_{0}(\cos\theta_{i},\sin\theta_{i})$, with $i=1,\,2\ldots\,N$ and $v_{0}$ a constant speed, on a square $L\times L$ periodic two-dimensional domain. At each time step each individual takes the average direction of those within some pre-defined radius $R$ as its new direction. Thus:
\begin{subequations}\label{eq:vicsek_model}
\begin{gather}
\bm{r}_{i}(t+\Delta t) = \bm{r}_{i}(t)+\bm{v}_{i}(t)\,\Delta t\;,\\[5pt]
\theta_{i}(t+\Delta t) = \langle \theta_{i}(t) \rangle_{R} + \xi_{i}\;,
\end{gather}
\end{subequations}
where $\bm{r}_{i}$ is the position of the $i-$th individual at time $t$, $\xi_{i}$ a uniformly distributed random angle in the range $[-\eta,\eta]$ and $\langle \theta_{i} \rangle_{R}=\arctan\langle \sin\theta_{i} \rangle_{R}/\langle \cos\theta_{i} \rangle_{R}$ is the average orientation of all the individuals within a distance $R$ from $\bm{r}_{i}$, including the $i-$th one. Following a classic convention, we set $\Delta t = R = 1$, thus choosing $\Delta t$ as unit of time and the interaction range $R$ as unit of distance. 

\begin{figure}[t]
\includegraphics[width = \columnwidth]{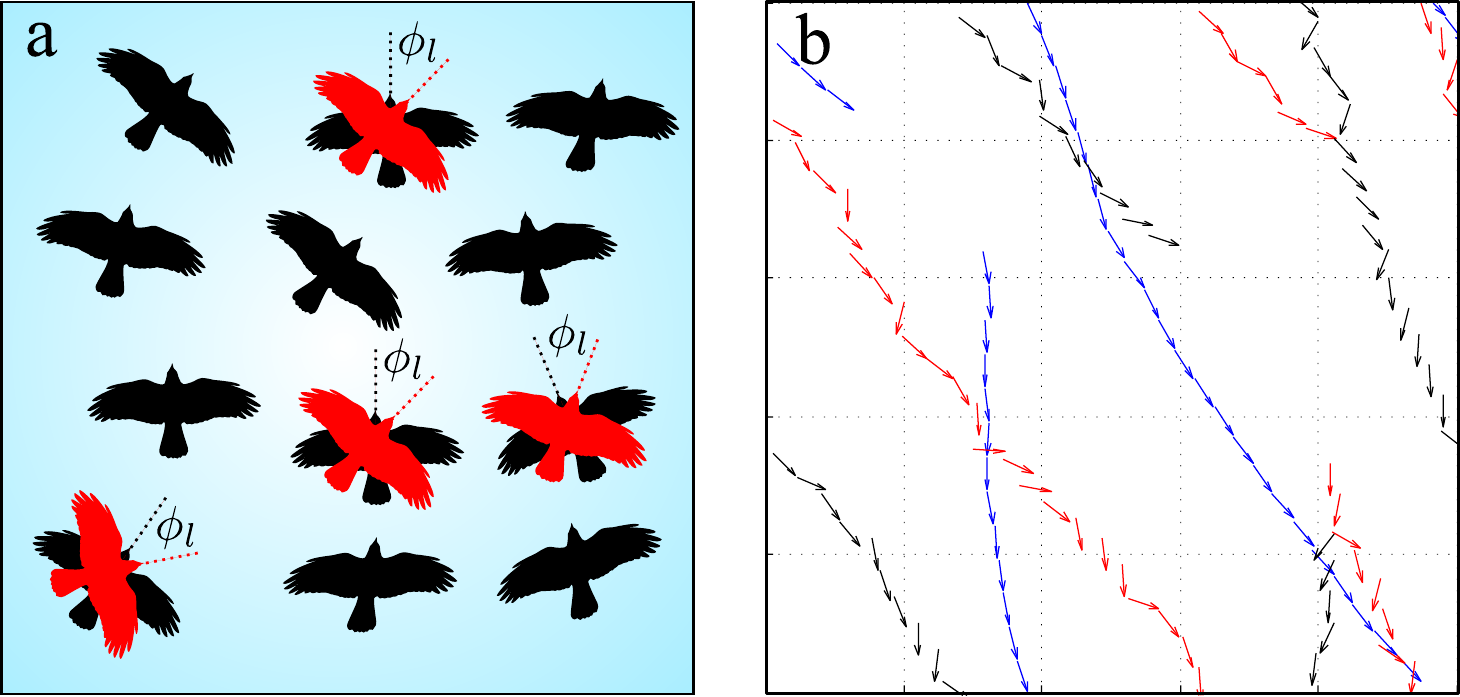}
\caption{\label{fig:snap} (a) Example of how the perturbation is applied to the flock, in this sketch $N_{l}=4$ leaders out of $N=12$ individuals turn by an angle $\phi_{l}$, thus changing their direction of motion from the black to the red dashed line. (b) Snapshot of a typical simulation; shown is the trajectory of one of the perturbed particles (red), a normal unperturbed particle (black) and a tracer particle inserted into the system that has $\eta = 0$ (blue). This represents 200 consecutive time steps from a simulation of $N = 1600$ particles with $\rho = 16$, $\eta = 0.25$, $N_{l} = 50$ and $\phi_{l} = 0.1$.}
\end{figure}

\begin{figure}[t]
\includegraphics[width = \columnwidth]{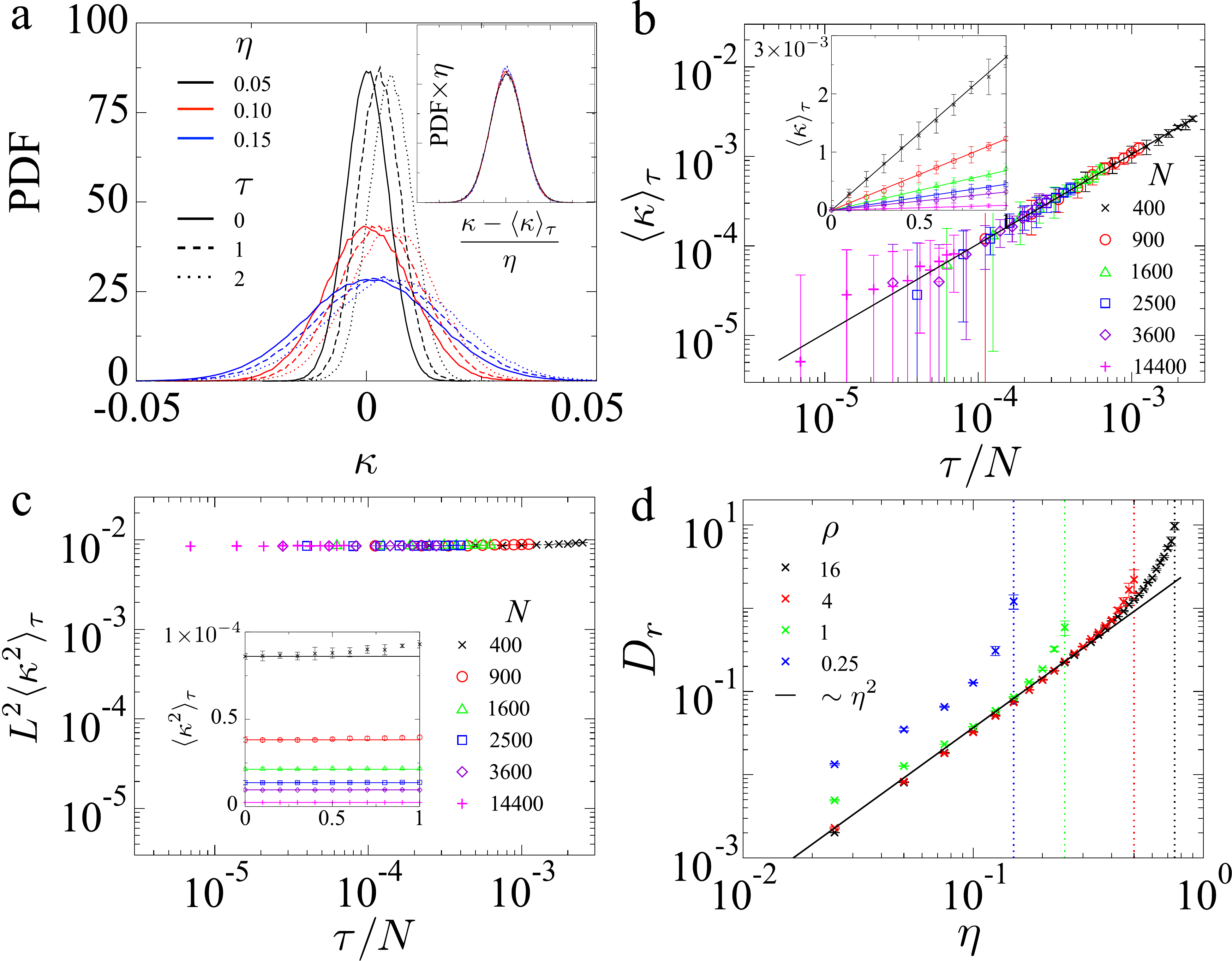}
\caption{\label{fig:linear_response} (a) Probability distribution function (PDF) of the discrete curvature $\kappa$ defined in Eq. \eqref{eq:curvature} for various effective torque and noise values. Inset: PDF of $(\kappa-\langle \kappa \rangle_{\tau})/\eta$ normalized by the noise standard deviation $\eta$. The data collapse on the same Gaussian. (b) The mean curvature $\langle\kappa\rangle_{\tau}$, of the trajectory of the flock is linear with the torque per individual $\tau/N$ (the black line shows $\langle\kappa\rangle_{\tau} \sim \tau/N$). (c) Conversely, the total mean-square curvature, $L^{2}\langle\kappa^2\rangle_{\tau}$, remains constant for small $\tau$ values. (d) The rotational diffusion coefficient $D_{r}$, defined in Eq. \eqref{eq:linear_response}, follows a universal power law relationship $D_{r}\sim\eta^{2}$ at high density.}
\end{figure}

In order to study the linear response of the flock to a perturbation, we consider the system polarization vector, defined as $\bm{P}(t)=1/(v_{0}N)\sum_{i=1}^{N}\bm{v}_{i}(t)$. The magnitude $P=|\bm{P}|$ serves as an order parameter and allows to distinguish the isotropic (where $P=0$) and flocking ($P > 0$) phase \cite{chate2008collective,ginelli2015physics}. The unit vector $\bm{p}=\bm{P}/P$, on the other hand, represents the global direction of the flock. Now, deeply in the order phase (i.e $P\sim 1$), $\bm{p}$ changes very slowly and the polarization vector randomly precesses along the unit circle (Supplementary Movie S1 \cite{SInote}). To quantify this process we introduce a discrete analog of curvature in the flock trajectory:
\begin{equation}\label{eq:curvature}
\kappa(t) = [\bm{p}(t-1)\times\bm{p}(t)]\cdot\bm{\hat{z}}\;.	
\end{equation}
In the absence of any rotational bias, the flock is equally likely to turn left or right, and $\kappa$ follows the Gaussian distribution (Fig. \ref{fig:linear_response}a) with zero mean value and finite variance. Next, let us consider a subset of $N_{l} \le N$ randomly chosen ``informed individuals'' within the flock, who are biased to turn toward a specific direction. For each of them, Eq. (\ref{eq:vicsek_model}b) is replaced by $\theta_{i}(t+\Delta t)=\phi_{l}+\langle \theta_{i}(t) \rangle_{R}+\xi_{i}$, where $\phi_{l}$ is a constant angular displacement representing the ``degree of influence'' of each informed individual within its neighborhood. While there are other ways to introduce an internal bias in the Vicsek model \cite{czirok1997spontaneously,chate2008collective}, this is possibly the one that most closely resembles maneuvers in real flocks. The product $\tau = \phi_{l}N_{l}$ is analogous to an \emph{effective torque} that is able to bend the trajectory of flock toward the left or right, depending on the sign of $\phi_{l}$ (see Fig.~\ref{fig:snap}b and (Supplementary Movie S2 \cite{SInote}).

In order to investigate the influence of the informed individuals in the general behavior of the flock, we have performed various numerical simulations (Fig.~\ref{fig:linear_response} and Appendix B). As a consequence of the directional bias introduced by the informed individuals, the trajectory of the flock acquires a non-zero mean curvature that grows linearly with the torque per individual: i.e. $\langle \kappa \rangle_{\tau}\sim \tau/N$, where $\langle \dots \rangle_{\tau}$ represent a time average in the presence of an effective torque $\tau$ (Fig, \ref{fig:linear_response}b). It is worth stressing that the linear response of the flock to leadership is not independently governed by the number of leaders $N_{l}$ or their influence $\phi_{l}$, but rather by their product $\tau$, so that doubling the number of leaders in the group is equivalent to keeping their number fixed, while doubling their influence (Fig.~\ref{fig:phiTemps}). The total mean squared curvature $L^{2}\langle \kappa^{2} \rangle_{\tau}$ is, on the other hand, independent of $\tau$ for small $\tau$ values and plateaus to the curvature variance $\langle \kappa^{2}\rangle_{0}$ of the unbiased flock (Fig, \ref{fig:linear_response}c). As a consequence, the ratio between the slope ${\rm d}\langle\kappa\rangle_{\tau}/{\rm d\tau}$ and $\langle\kappa^{2}\rangle_{0}$ depends on the flock population and size only through its density $\rho=N/L^{2}$. This allows us to formulate the following linear response relation for the Vicsek model subject to the leadership of a subset of informed individuals:
\begin{equation}\label{eq:linear_response}
\frac{{\rm d}\langle \kappa \rangle_{\tau}}{{\rm d}\tau} = \frac{1}{2D_{r}}\,\langle \kappa^{2} \rangle_{0}\;.
\end{equation}
where $D_{r}$ is an effective rotational diffusion coefficient, generally dependent on the system density, noise and particle velocity. In order to gain insight into the dependence of $D_{r}$ on the remaining free parameters of the system, we have repeated the previous analysis for various $v_0$, $\rho$ and $\eta$ values (Fig. \ref{fig:linear_response}d and Fig.~\ref{fig:V0comparison}) and find that, surprisingly, the density and velocity dependence disappears at high densities, revealing a universal linear dependence of $D_{r}$ on the variance of noise: $D_{r}\sim \eta^{2}$.

Some comments are in order. Eq. \eqref{eq:linear_response} is a special case of the fluctuation-dissipation theorem (FDT), with a collective effective temperature $T_{\rm eff}\sim D_{r}$. In the realm of active matter, the possibility of an effective temperature and a generalized FDT has been discussed for both dilute and dense phases \cite{czirok1997spontaneously,chate2008collective,loi2008effective,wang2011communication,loi2011non-conservative,berthier2013non-equilibrium,szamel2014self-propelled,levis2015single}, sometimes with contradictory results. {Unlike in equilibrium systems, where any perturbation $\Delta\mathcal{H}=-fB$ to the Hamiltonian results into a fluctuation-dissipation relation of the form:
\begin{equation}\label{eq:equilibrium_fdt}
\frac{{\rm d}\langle A \rangle_{f}}{{\rm d}f} = \frac{1}{k_{B}T}\,\left[\langle AB \rangle_{0}-\langle A \rangle_{0} \langle B \rangle_{0}\right]\;,
\end{equation}
where $A$ and $B$ are generic observables, $f$ is an external field coupled to the observable $B$ and and $T$ is a {\em unique} temperature, relations like Eq. \eqref{eq:equilibrium_fdt} hold, outside of equilibrium, only for specific choice of observables and perturbations}. Czir\'ok {\em et al.} \cite{czirok1997spontaneously} analyzed, for instance, the response of a Vicsek flock to a spatially uniform aligning field and found no evidence of a fluctuation-dissipation relation. This was instead identified by Chat\'e and coworkers \cite{chate2008collective}, who considered an external field coupled with the local average polarization. The effective temperature resulting from this relation, however, varies in the parameter space. {Our findings, indicate that a special form of the FDT, with an effective temperature only dependent on the variance of angular noise, can be identified in the Vicsek model subject to the leadership of a subset of informed individuals, as long as the system is sufficiently dense. We further stress that the linear response to nonuniform internal torques, discussed here, appears to be fundamentally different from the highly nonlinear response to uniform aligning fields recently reported by Kyriakopoulos {\em et al}. \cite{Kyriakopoulos2016leading}}.

\subsection{\label{sec:continuous_flocks}{Mean-field} continuous flocks}

To shed light on the numerical results presented in Sec. \ref{sec:discrete_flocks} and, in particular, the universal behavior of the coefficient $D_{r}$ at large densities, we have considered the effect of leadership in a {mean-field} ``continuous flock'' confined in a periodic domain. Calling $\rho$, $\bm{v}$ and $\bm{V}=\rho\bm{v}$, the flock density, velocity and momentum density, respectively, the flock dynamics is governed by a modified version of Toner-Tu hydrodynamics equations \cite{Toner1995Long}:
\begin{subequations}\label{eq:toner_tu}
\begin{align}
&\partial_{t}\rho + \nabla\cdot\bm{V} = 0\;, \\
&\partial_{t}\bm{V}+\lambda_{1}(\bm{V}\cdot\nabla)\bm{V}+\lambda_{2}(\nabla\cdot\bm{V})\bm{V}+\lambda_{3}\nabla|\bm{V}|^{2} \notag\\
&= \bm{\Omega}\times\bm{V} + (\alpha-\beta|\bm{V}|^{2})\bm{V}-\nabla \Pi + \nu\nabla^{2}\bm{V}+\bm{f}\;,
\end{align}
\end{subequations}
where $\Pi$ is a density dependent pressure and $\bm{f}$ is a delta-correlated random force, such that: $\langle f_{i}(\bm{r},t)f_{j}(\bm{r}',t')\rangle = 2D \delta_{ij}\delta(\bm{r}-\bm{r}')\delta(t-t')$. {Interestingly, equations very similar to these have been also used by Alicea {\em et al}. to explain zero resistance states in two-dimensional electron electron gasses driven by microwaves \cite{alicea2005transition}.} The amplitude $D$ is evidently proportional to the variance of the angular noise in the original Vicsek model, i.e. $D\sim\eta^{2}$. The first term on the right-hand side of Eq. (\ref{eq:toner_tu}b), representing a uniform rigid body rotation with constant angular velocity $\bm{\Omega}=\Omega\bm{\hat{z}}$, embodies the effect of the torque exerted by the informed individuals in the flock. In order to recover the high-density limit, we assume the system to be incompressible and strongly polarized. {The former assumption, in particular, has profound consequences on the critical behavior of the system and, as it was recently demonstrated in Refs.\cite{checn2015critical,chen2016mapping}, ascribes the continuous (incompressible) and discrete (compressible) model to different universality classes. As we will see later, this distinction, however, does not affect the response to leadership, as long as the system is deeply in the polarized state, and thus well below the flocking phase transition.} After some algebraic manipulations (see Appendix A), Eqs. \eqref{eq:toner_tu} yield the following equation for the total polarization:
\begin{equation}\label{eq:p_dynamics}
\partial_{t}\bm{P} = \bm{\Omega}\times\bm{P}-2\alpha\delta P\,\bm{p}+\bm{F}+O(\delta P^{2})\;.	
\end{equation}
where $\delta P$ is the departure of the order polarization amplitude from its mean value $P_{0}=\sqrt{\alpha/\beta}$ and $\bm{F}=\int {\rm d}A\,\bm{f}$ is a Gaussianly distributed random function delta-correlated in time: $\langle F_{i}(t)F_{j}(t) \rangle = 2D\delta(t-t')$. According to Eq. \eqref{eq:p_dynamics}, the dynamics of the total polarization depends on the various parameters appearing in Eq. \eqref{eq:toner_tu}, exclusively through the fluctuations of the order parameter. Thus, consistently with our numerical simulations, the total polarization approaches a universal behavior as the flock becomes uniformly polarized. In this regime, $\delta P$ decays exponentially over a time scale of order $1/\alpha$. Thus, for $\Omega \ll \alpha$, projecting Eq. \eqref{eq:p_dynamics} on the transverse direction yields:
\begin{equation}\label{eq:rotational_diffusion}
\partial_{t}\bm{p} = \bm{\Omega}\times\bm{p}+\bm{F}_{\perp}/P_{0}\;,
\end{equation}
where $\bm{F}_{\perp}=\bm{F}-(\bm{p}\cdot\bm{F})\bm{p}$. The discrete curvature $\kappa$ can be obtained by integrating the local curvature over a finite time interval $\Delta t$, namely:
\begin{equation}\label{eq:continuous_curvature}
\kappa(t) = \int_{t-\Delta t}^{t}{\rm d}t'\,(\bm{p}\times\partial_{t'}\bm{p})\cdot\bm{\hat{z}}\;.
\end{equation}
As a consequence of Eqs. \eqref{eq:rotational_diffusion} and \eqref{eq:continuous_curvature} $\kappa$ is Gaussianly distributed, with $\langle \kappa \rangle_{\tau}=\Omega\Delta t$ and $\langle \kappa^{2} \rangle_{\tau}=(\Omega \Delta t)^{2}+2D\Delta t/P_{0}^{2}$. This, finally, implies the linear response relation \eqref{eq:linear_response}, with $\tau=\Omega\Delta t$ and $D_{r} = D\Delta t/P_{0}^{2}\sim \eta^{2}$. In order to verify the quality of the agreement between our discrete and continuous model, we have plotted the normalized probability distribution function of $(\kappa-\tau)/\eta$ ($P_{0}=1$ in our simulations) for various torques and noise variances (Fig. \ref{fig:linear_response}a inset). As predicted by our continuous model, the data fall on the same Gaussian.

\begin{figure}[t]
\includegraphics[width = \columnwidth]{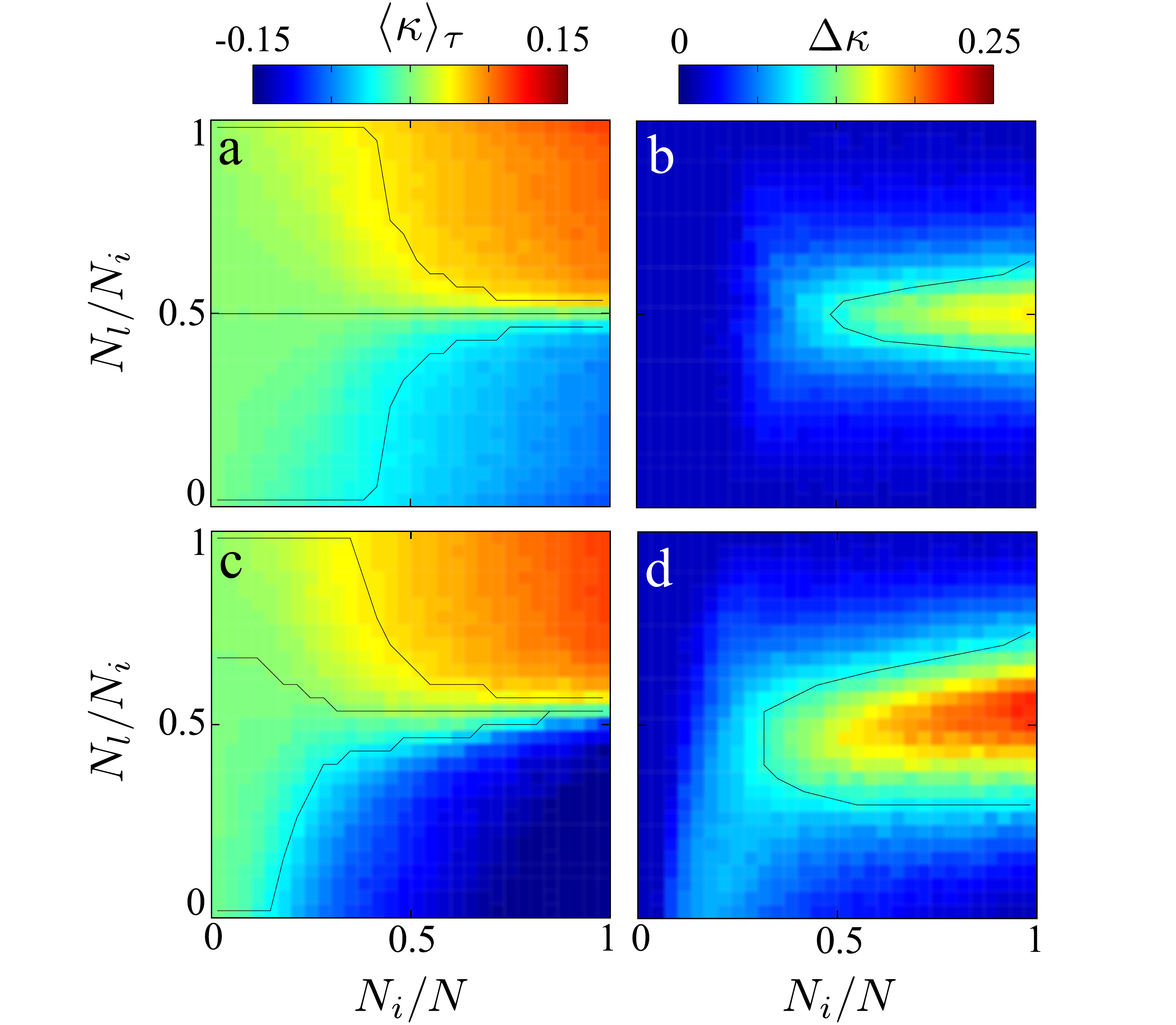}
\caption{\label{fig:decision_making} Induced mean (a,c) and standard deviation (b,d) curvature in the motion of the flock due to competing subsets of informed individuals. Here $N_l$ and $N_r$ individuals giving the flock a positive and negative curvature respectively, with $N_i=N_l+N_r$. (a,b) When the competing subsets have equal and opposite influence ($\phi_l=-\phi_r$) the resulting curvature is of the same sign as the largest informed subset, and the inclusion of uninformed individuals reduces the standard deviation of the curvature. (c,d) When $\phi_r=-2\phi_l$ the resulting curvature is distorted and the relative sizes of the informed subsets is no longer sufficient to predict the curvature. The black lines in (a,c) show $\langle\kappa\rangle_{\tau} = -0.04, 0, +0.04$ and in (b,d) show $\langle\kappa^2 \rangle_{\tau} = 0.1$, all simulations here are done with $N=900$, $\rho=4$, $v_0=0.1$, $\phi_l=0.1$.}
\end{figure}

\subsection{\label{sec:decision_making}Decision making}

We next turn our attention to how flocks make decisions. As mentioned earlier, combined experimental and theoretical studies on schooling fish \cite{couzin2005effective,ward2008quorum,sumpter2008consensus,couzin2011uninformed} have demonstrated that, in the presence of competing interests (i.e. such as swimming toward two different targets), the group decides to conform to the behavior of the largest minority with an accuracy that increases with the number of uninformed individuals. Our approach allows us to study this result in a system where the response can vary continuously. To this purpose, we have introduced a second subset of $N_{r}$ individuals with an angular displacement $\phi_r=-\phi_l$, so that $N_{i} = N_{l} + N_{r}$ is now the total number of informed individuals and the flock must decide between two competing informed subsets. Fig.~\ref{fig:decision_making}a shows that the sign of the resulting curvature of the flock is dictated by the largest subset of informed individuals; this is true even when the competing subsets in the flock are of similar size, $N_l\approx N_r$. However, as the number of informed individuals becomes large, $N_i \approx N$, the standard deviation in the curvature, $\Delta\kappa = \sqrt{\langle \kappa^{2}\rangle_{\tau}-\langle \kappa \rangle_{\tau}^{2}}$, significantly increases and the flock becomes less efficient at selecting the correct behavior. This is because when most individuals in the flock are leaders, with either a positive or negative curvature, there are few followers to average out the competing effects. Fig.~\ref{fig:decision_making}c shows the response of a flock when $\phi_{r} = -2\phi_{l}$. Here we see again that the presence of uninformed individuals reduces the standard deviation, but the resulting curvature is no longer symmetric around $N_{l}=N_{r}$. Hence a smaller but sufficiently influential subgroup can dictate the sign of the curvature of the flock trajectory.

\begin{figure}[t]
\includegraphics[width = \columnwidth]{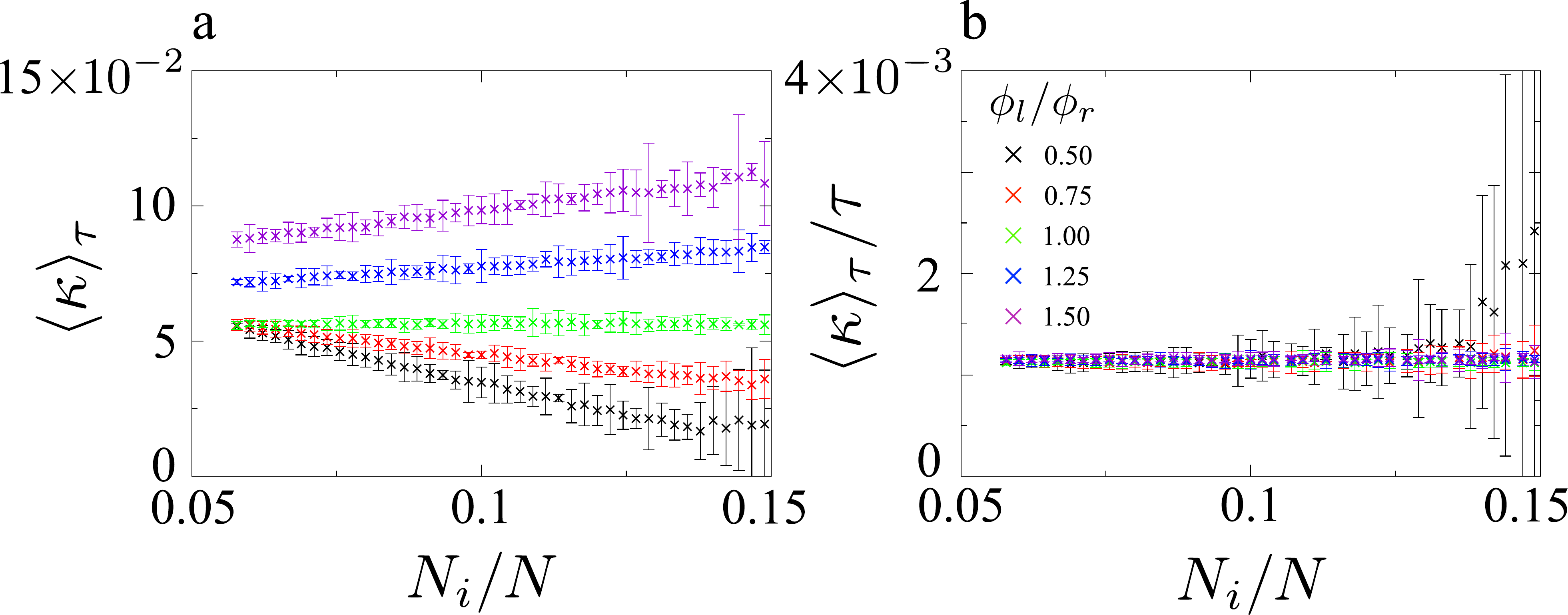}
\caption{\label{fig:decision_making_torque}a) Resulting curvature of flocks with $N_l-N_r$ fixed and varying values of $\phi_r$. As is clear here the resulting curvature is not just the function of the relative sizes of the informed subsets, rather it is the total torque. b) The ratio between the resulting curvature and the total torque is constant across all simulations, indicating this is the only important parameter here. All simulations here are done with $N=900$, $\rho=4$, $v_0=0.1$, $\phi_l=0.1$, $N_l-N_r=50$.}
\end{figure}

Our results suggest that linear response to leadership can be extended in the presence of two competing subsets of informed individuals upon introducing a generalized effective torque $\tau = \phi_{l}N_{l}+\phi_{r}N_{r}$. To verify this we fixed the relative size of the informed subsets ($N_r-N_l$) and vary their relative influence $\phi_l/\phi_r$ achieving a range of responses, see Fig.~\ref{fig:decision_making_torque}a. By normalizing the response curvature by this newly defined $\tau$ we find that all the results fall onto the same value, Fig.~\ref{fig:decision_making_torque}b. This confirms that the linear response to leadership extends to competing subgroups of leaders implying that all these flocks are at the same effective temperature. Hence flocks do not merely select the behavior of the largest subset of informed individuals, but rather the resulting behavior is the response to the total influence of both subsets on the flocks (the same result can be achieved by fixing $\phi_l/\phi_r$ and varying $N_r-N_l$, see Fig.~\ref{fig:phiTemps}).

\begin{figure*}[t]
\includegraphics[width=2\columnwidth]{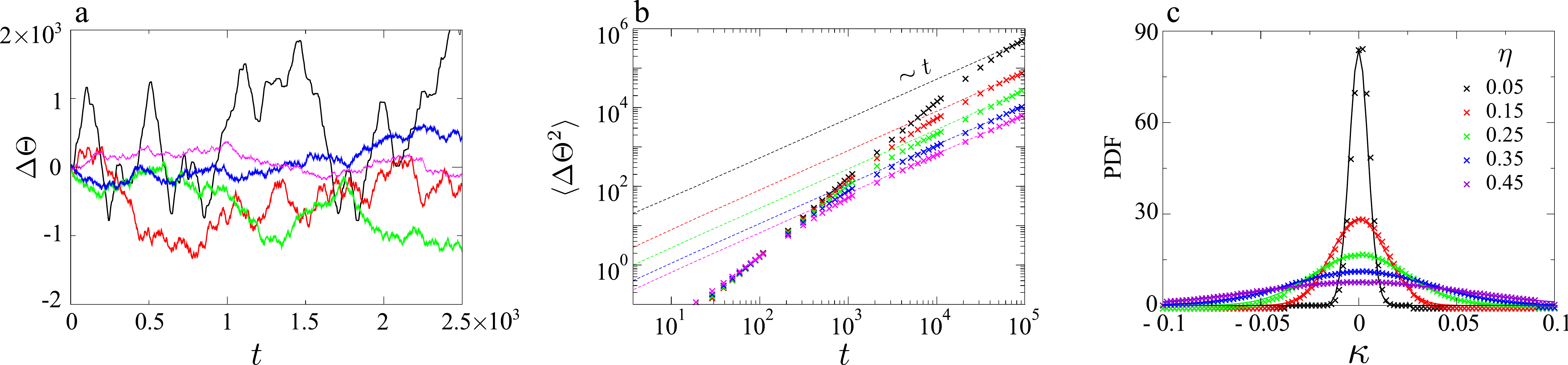}
\caption{\label{fig:Diffusive} (a) Flock angular displacement $\Delta\Theta$, with $\Theta$ defined from $\bm{p}=(\cos\Theta,\sin\Theta)$, as a function of time. (b) Mean squared angular displacement versus time illustrating the long-time diffusive behavior. (c) Probability distribution function (PDF) for the discrete curvature $\kappa$. All results here are for systems of $N=400$ particles with $\rho=4$ and the data were taken over $320 \times 10^{4}$ time steps after a $150 \times 10^{3}$ step relaxation period.}
\end{figure*}

\section{Discussion and Conclusions}

We have investigated how a model flock respond to the leadership of a subset of informed individuals. Using numerical simulations and continuous modeling, we have demonstrated that the process obeys to a special form of the fluctuation dissipation theorem (FDT), with an emerging effective temperature that depends uniquely on the variance of noise for sufficiently dense systems. The linear response to leadership also holds in the presence of two subgroups of informed individuals with competing interests, in this case the flock behavioral decision is determined by both the number of leaders and their degree of influence. So that a small subgroup of particularly influential informed individuals can overrule a larger subset of less influential informed individuals. Our theoretical results provide general insight into leadership in collective behavior and can be used to shed light on previous experimental studies, particularly those on schooling fish \cite{ward2008quorum,sumpter2008consensus,couzin2011uninformed}. In bird flocks, recent observations have demonstrated that maneuvers can be initiated by a single leader \cite{attanasi2014information,attanasi2015emergence} with the reaction of the nearby birds propagating linearly through the flock. This spatial and structural dependence of the recruitment of leaders, and subsequent followers, is a clear next step for the work presented here.

\begin{acknowledgments}
We would like to thank Matthew Turner and Denis Bartolo for helpful discussions while producing this work. This work was supported by the Netherlands Organization for Scientific Research (NWO/OCW), as part of the Frontiers of Nanoscience program. 
\end{acknowledgments}


\appendix

\section{\label{sec:appendix_continuous_model}Derivation of mean-field equations for continuous flocks}

In this section we provide a derivation of Eqs. \eqref{eq:p_dynamics} and \eqref{eq:rotational_diffusion} in the Sec. \ref{sec:continuous_flocks} as a more detailed discussion of our continuous model. The main object of the discrete model is the discrete curvature defined as:
\begin{equation}
\kappa(t) = [\bm{p}(t-\Delta t)\times\bm{p}(t)]\cdot\bm{\hat{z}}\;,
\end{equation}
where $\bm{p}$ is the average flock direction calculated from the macroscopic polarization $\bm{P}=P\bm{p}$, with $P$ the polar order parameter. In practice, the quantity $\kappa(t)$ represents the global angular displacement experienced by the flock in the time interval $[t-\Delta t,t]$, with $\Delta t=1$ in the units used in our numerical simulations.  

\begin{figure*}[t]
\includegraphics[width=2\columnwidth]{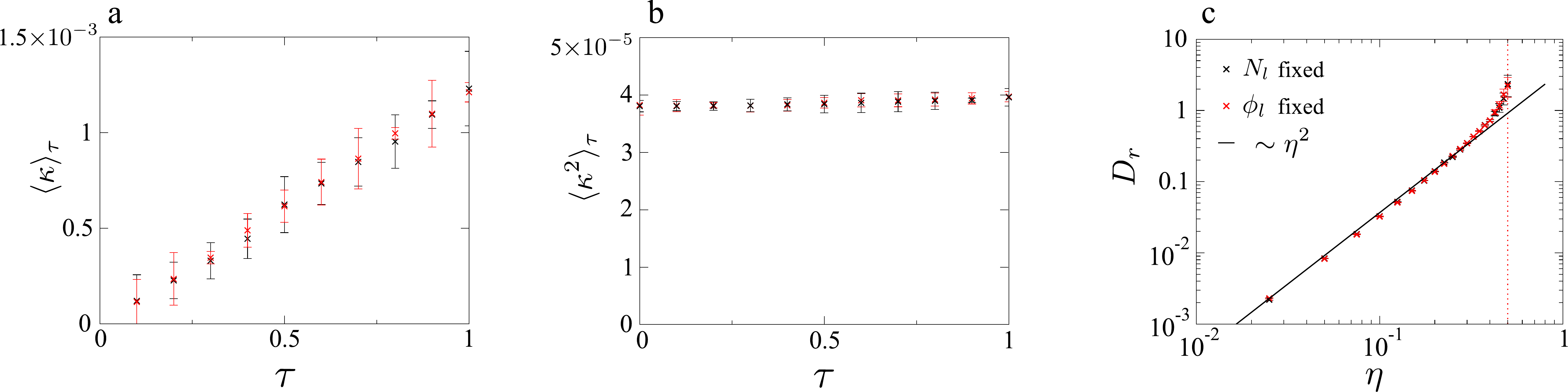}
\caption{\label{fig:phiTemps} Mean curvature (a) and mean squared curvature (b) versus torque. (c) Rotational diffusion coefficient versus noise. All these quantities depend exclusively on the torque, $\tau=\phi_{l}N_{l}$, regardless of whether it is adjusted by varying the number of leaders $N_{l}$ or their degree of influence $\phi_{l}$. (a) and (b) both show the results for $N=900$ and $\eta = 0.1$. The black and red points represent simulations where $N_l = 10$ and $\phi_l = 0.1$, respectively. All results here are for particles with $\rho=4$, $v_0=0.1$ and the data was taken over $150 \times 10^{3}$ time steps after a $10^{4}$ step relaxation period.}
\end{figure*}

\begin{figure*}
\includegraphics[width=2\columnwidth]{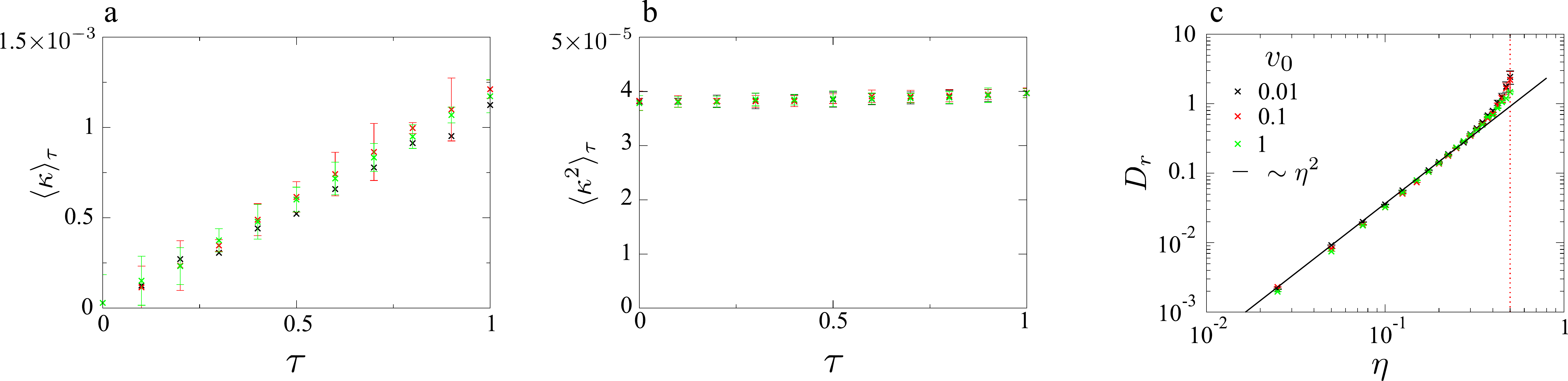}
\caption{\label{fig:V0comparison} Mean curvature (a) and mean squared curvature (b) versus torque. (c) Rotational diffusion coefficient versus noise. All these quantities are independent on the particle speed $v_0$. (a) and (b) both show the results for $N=900$ and $\eta = 0.1$. All results here are for systems with $\rho=4$ and the data was taken over $150 \times 10^{3}$ time steps after a $10^{4}$ step relaxation period.}
\end{figure*}

Let us then consider a ``continuous flock'' confined in a periodic domain. The system is described by a local density $\rho$ and velocity field $\bm{v}$, as well as the local momentum density $\bm{V}=\rho\bm{v}$. The macroscopic polarization  is then defined as:
\begin{equation}
\bm{P} = \int {\rm d}A\,\bm{V}\;.	
\end{equation}
Analogously, a continuous version of $\kappa$ can be constructed as follows:
\begin{equation}
\kappa(t) = \int_{t-\Delta t}^{t} {\rm d}t'\,(\bm{p}\times\partial_{t'}\bm{p})\cdot\bm{\hat{z}}\;.
\end{equation}
Now, within the classic hydrodynamic framework, first introduced by Toner and Tu \cite{Toner1995Long}, the dynamics of the density and momentum density are governed, in the absence of leadership, by the following set of stochastic partial differential equations:
\begin{subequations}\label{eq:toner_tuSI}
\begin{gather}
\begin{align}
&\partial_{t}\rho + \nabla\cdot\bm{V} = 0\;,\\[5pt]
&\partial_{t}\bm{V}+\lambda_{1}(\bm{V}\cdot\nabla)\bm{V}+\lambda_{2}(\nabla\cdot\bm{V})\bm{V}+\lambda_{3}\nabla|\bm{V}|^{2}\notag \\
&= (\alpha-\beta |\bm{V}|^{2})\bm{V}-\nabla \Pi + \nu \nabla^{2} \bm{V} + \bm{f}\;.
\end{align}
\end{gather}	
\end{subequations}
where $\Pi$ is a density dependent pressure and $\bm{f}$ a delta-correlated random force:
\begin{equation}
\langle f_{i}(\bm{r},t)f_{j}(\bm{r}',t') \rangle = 2D \delta_{ij} \delta(\bm{r}-\bm{r}')\delta(t-t')\;,
\end{equation}
with $D$ proportional to the variance of the angular noise in the original Vicsek model: $D \sim \eta^{2}$. In order to recover the high density limit, we assume the system to be incompressible and strongly polarized. Using the first of these assumptions yields $\nabla\cdot\bm{V}=0$, while Eq. (\ref{eq:toner_tuSI}b) simplifies to the form:
\begin{equation}\label{eq:toner_tu_simplified} 
\partial_{t}\bm{V} = (\alpha-\beta|\bm{V}|^{2})\bm{V}+\nabla\cdot\bm{\Sigma}+\bm{f}\;,	
\end{equation}
where $\bm{\Sigma}$ is an effective stress tensor given by:
\begin{equation}
\Sigma_{ij} = -(\Pi+\lambda_{3}|\bm{V}|^{2})\delta_{ij} - \lambda_{1}V_{i}V_{j} + \nu (\partial_{i}V_{j}+\partial_{j}V_{i})\;.
\end{equation}
Integrating Eq. \eqref{eq:toner_tu_simplified} in space yields then a dynamical equation for the macroscopic polarization $\bm{P}$. Namely:
\begin{equation}\label{eq:dot_P_1}
\partial_{t}\bm{P} = \int {\rm d}A\,\partial_{t}\bm{V} = \bm{F}-\int {\rm d}A\,(\alpha-\beta|\bm{V}|^{2})\bm{V}\;,
\end{equation}
where $\bm{F}=\int {\rm d}A\,\bm{f}$ and we used the fact that $\int {\rm d}A\,\nabla\cdot\bm{\Sigma}=0$, from the divergence theorem on a periodic domain. By virtue of the central limit theorem, the function $\bm{F}$ is Gaussianly distributed, so that:
\begin{equation}
\langle F_{i}(t)F_{j}(t') \rangle = 2D\,\delta_{ij}\delta(t-t')\;.
\end{equation} 
Now, if the flock is strongly polarized, $\bm{V}=\sqrt{\alpha/\beta}\,\bm{p}+\delta\bm{V}$, with $|\delta\bm{V}|\ll 1$. Then, expanding Eq. \eqref{eq:dot_P_1} at the linear order gives:
\begin{equation}\label{eq:dot_P_2}
\partial_{t}\bm{P} = \bm{F}-2\alpha (\bm{p}\cdot\delta\bm{P})\bm{p}+O(|\delta\bm{P}|^{2})\;,
\end{equation}
where $\delta\bm{P}=\int {\rm d}A\,\delta\bm{V}$. The quantity $\bm{p}\cdot\delta\bm{P}$ represents, at the linear order, the fluctuations in the order parameter, since $\delta\bm{P}=\delta P \bm{p}+P\delta\bm{p}$ and $\bm{p}\cdot\delta\bm{p}=0$, being $\bm{p}$ a unit vector. Now, deeply in the order phase and for large positive $\alpha$, the fluctuations of the order parameter decay exponentially in time (i.e. $\langle \delta P \rangle \sim e^{-2\alpha t}$). Thus, assuming that $\delta P$ has relaxed to zero so that $P=\sqrt{\alpha/\beta}=P_{0}$, and projecting Eq. \eqref{eq:dot_P_2} on the transverse $\bm{p}-$direction, we obtain:
\begin{equation}
\partial_{t}\bm{p} = \bm{F}_{\perp}/P_{0}\;,	
\end{equation}
where $\bm{F}_{\perp}$ is the transverse component of $\bm{F}$. From this the curvature can be readily calculated:
\begin{equation}
\kappa(t) = \int_{t-\Delta t}^{t} {\rm d}t'\,F_{\perp}\;.
\end{equation}
Thus, in the absence of leadership, the curvature $\kappa(t)$ is a Gaussianly distributed random number having zero mean and variance:
\begin{equation}\label{eq:kappa_0}
\langle \kappa^{2} \rangle_{0} = \frac{2D \Delta t}{P_{0}^{2}};,
\end{equation}
consistently with our numerical data (Fig.~\ref{fig:linear_response}b and Fig.~\ref{fig:Diffusive}c). Now, leadership can be incorporated in this model by introducing a uniform rotation, with angular velocity $\bm{\Omega}=\Omega\bm{\hat{z}}$, into Eq. \eqref{eq:toner_tuSI}. In the incompressible limit, this yields:
\begin{equation}
\partial_{t}\bm{V} = \bm{\Omega}\times\bm{V}+(\alpha-\beta|\bm{V}|^{2})\bm{V}+\nabla\cdot\bm{\Sigma}+\bm{f}\;.
\end{equation}
Notice that, by virtue of the definition of $\bm{V}$, the performance of such a mechanism depends on both the system density and orientational order, consistently with our numerical model. Proceeding as before, one straightforwardly gets:
\begin{equation}
\partial_{t}\bm{p} = \bm{\Omega}\,\times\bm{p}+\bm{F}_{\perp}/P_{0} 	
\end{equation} 
from which:
\begin{equation}\label{eq:kappa_omega}
\langle \kappa \rangle_{\tau} = \Omega\,\Delta t\;, \qquad	
\langle \kappa^{2}\rangle_{\tau} = (\Omega\,\Delta t)^{2}+\frac{2D\Delta t}{P_{0}^{2}}\; \qquad	
\end{equation}
Eqs. \eqref{eq:kappa_0} and \eqref{eq:kappa_omega} finally imply the linear response relation given in Eq.~\eqref{eq:linear_response} with $\tau=\Omega\,\Delta t$ and $D_{r}=D\Delta t/P_{0}^{2}\sim \eta^{2}$.

\section{Supplementary Numerical Data}

\begin{figure}
\includegraphics[width=\columnwidth]{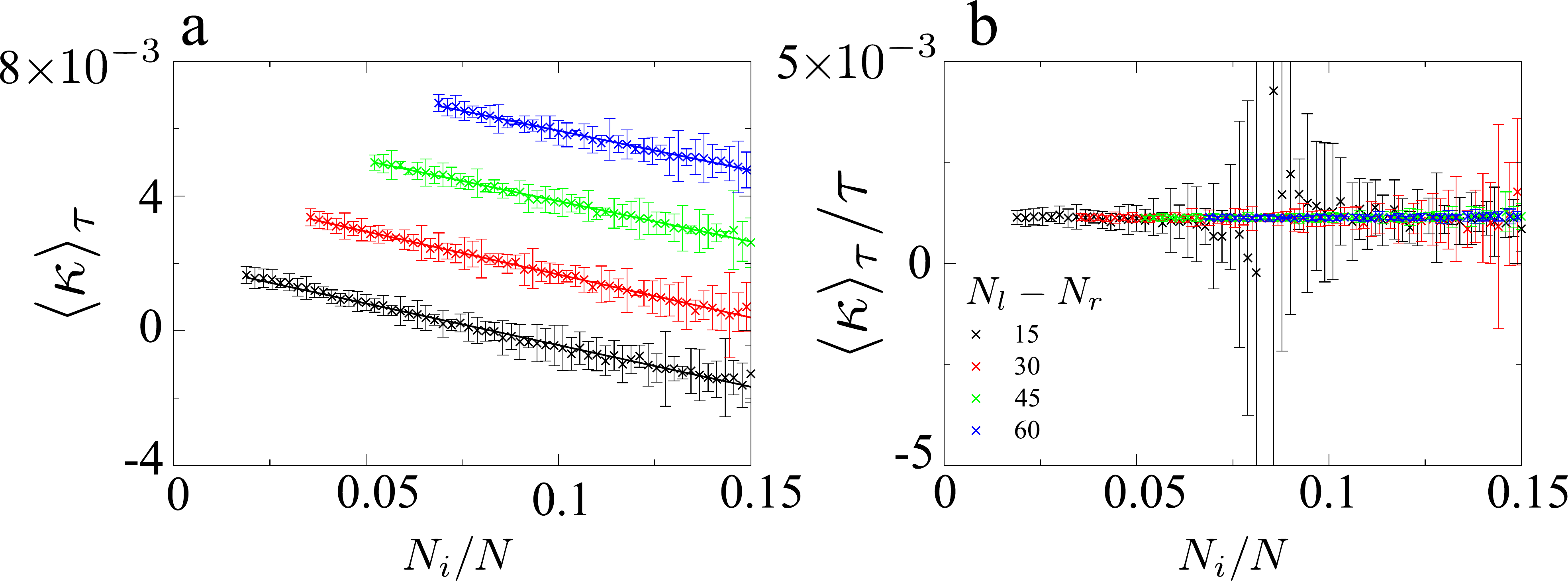}
\caption{\label{fig:competing} Mean curvature (a) and mean squared curvature (b) in the presence of two competing subsets of leaders, $N_l$ with influence $\phi_l$ and $N_r$ with influence $\phi_r$, here we have set $\phi_r = -1.5\phi_l$. When the curvature is normalized by the total torque applied to the system all simulations give the same response, hence the flock is acting like a thermal bath with a linear response to leadership. All results here are for systems with $N=900$, $v_0=0.1$, $\rho=4$ and the data was taken over $150 \times 10^{3}$ time steps after a $10^{4}$ step relaxation period.}
\end{figure}

\subsection{Numerical Methods}

All simulations and analysis were performed using code written by the D. J. G. Pearce in C++. The code follows a slightly modified version of the original Vicsek model outlined in Ref.~\cite{vicsek}. All simulations were pre-equilibrated by a minimum of $10^{4}$ time steps, significantly longer than the autocorrelation time in the curvature of an unperturbed flock. 

Each point in Fig.~\ref{fig:linear_response}b,c corresponds to an average calculated over simulation of $150 \times 10^{3}$ time steps with $v_0=0.1$, $\rho=N/L^2=4$, $\eta=0.1$. Sixty Six such regressions were then created in order to create each point in Fig.~\ref{fig:linear_response}d. This is also true for all points represented in Figs.~\ref{fig:phiTemps} and \ref{fig:V0comparison}. Each point on the color plots in Fig.~\ref{fig:decision_making} corresponds to an average over a simulation of $150 \times 10^{3}$ time steps with $29$ increments in the $y-$direction and $30$ increments in the $x-$direction. The parameters used were $v_0=0.1$, $\rho=N/L^2=4$, $N=900$, $\eta=0.2$. Each point in Fig.~\ref{fig:decision_making_torque} corresponds to an average over a simulation of $150 \times 10^{3}$ time steps with $v_0=0.1$, $\rho=N/L^2=4$, $N=900$, $\eta=0.2$. Unless stated otherwise, all simulations are performed at $v_0=0.1$, $\rho=N/L^2=4$.

\subsection{Diffusive dynamics of the polarization direction}

To confirm the diffusive dynamics of the polarization direction $\bm{p}=(\cos\Theta,\sin\Theta)$ we simulated flocks at varying values of noise and studied the dynamics of the angular displacement $\Delta\Theta(t)=\Theta(t)-\Theta(0)$. Fig. \ref{fig:Diffusive}a,b demonstrate that the polarization direction $\bm{p}$ undergoes a random walk along the unit circle, so that $\langle \Delta\Theta(t)\rangle=0$ and $\langle \Delta\Theta(t)^{2}\rangle \sim t$ for $t\gg 0$. The corresponding discrete curvature $\kappa$ is Gaussianly distributed (Fig. \ref{fig:Diffusive}c).

\subsection{Dependence on the torque $\tau=\phi_{l}N_{l}$}

When adjusting the total applied perturbation in Fig.~\ref{fig:linear_response} we adjust the number of particles that the perturbation is applied to, $N_{l}$, while keeping $\phi_{l}$, constant. Fig.~\ref{fig:phiTemps} clearly shows that this is equivalent to adjusting the size of the perturbation by keeping $N_{l}$ constant and varying $\phi_{l}$. The flock response to leadership is then controlled by the effective torque $\tau=\phi_{l}N_{l}$.

\subsection{Dependence on the particle speed $v_{0}$}

The observed rotational diffusion coefficient does not depend on the particle speed $v_{0}$, Fig. \ref{fig:V0comparison}. This may break down when $v_{0} \gg  1$, and the arrangement of the flock is effectively randomized between time steps, or for $v_0 \sim 0$ and the arrangement becomes effectively fixed at accessible timescales. Here we choose to stay within the regime that recreates flocking like behavior similar to the collective motion of animals. 

\subsection{Dependence on the relative size of informed subsets}

Fig.~\ref{fig:decision_making_torque} shows that the response of a flock to the influence of two competing subsets is linear with the combined perturbation of the competing subsets. The relative magnitude of perturbation due to each of the two sets can be varied by adjusting the ratio $\phi_{l}/\phi_{r}$ (shown in Fig.~\ref{fig:decision_making_torque}), or by adjusting their relative size $N_{l}-N_{r}$. Fig. \ref{fig:competing} mirrors the analysis performed in Fig.~\ref{fig:decision_making_torque}, but now fixing $\phi_{l}/\phi_{r}$ and adjusting $N_{l}-N_{r}$ to recreate the same result. Additionally we see here that the result remains true even when $\tau$ changes sign, this crossover corresponds to the regions of Fig.~\ref{fig:competing}b where the normalized curvature appears to diverge since $\tau$ is very small.


\begin{thebibliography}{46}%
\makeatletter
\providecommand \@ifxundefined [1]{%
 \@ifx{#1\undefined}
}%
\providecommand \@ifnum [1]{%
 \ifnum #1\expandafter \@firstoftwo
 \else \expandafter \@secondoftwo
 \fi
}%
\providecommand \@ifx [1]{%
 \ifx #1\expandafter \@firstoftwo
 \else \expandafter \@secondoftwo
 \fi
}%
\providecommand \natexlab [1]{#1}%
\providecommand \enquote  [1]{``#1''}%
\providecommand \bibnamefont  [1]{#1}%
\providecommand \bibfnamefont [1]{#1}%
\providecommand \citenamefont [1]{#1}%
\providecommand \href@noop [0]{\@secondoftwo}%
\providecommand \href [0]{\begingroup \@sanitize@url \@href}%
\providecommand \@href[1]{\@@startlink{#1}\@@href}%
\providecommand \@@href[1]{\endgroup#1\@@endlink}%
\providecommand \@sanitize@url [0]{\catcode `\\12\catcode `\$12\catcode
  `\&12\catcode `\#12\catcode `\^12\catcode `\_12\catcode `\%12\relax}%
\providecommand \@@startlink[1]{}%
\providecommand \@@endlink[0]{}%
\providecommand \url  [0]{\begingroup\@sanitize@url \@url }%
\providecommand \@url [1]{\endgroup\@href {#1}{\urlprefix }}%
\providecommand \urlprefix  [0]{URL }%
\providecommand \Eprint [0]{\href }%
\providecommand \doibase [0]{http://dx.doi.org/}%
\providecommand \selectlanguage [0]{\@gobble}%
\providecommand \bibinfo  [0]{\@secondoftwo}%
\providecommand \bibfield  [0]{\@secondoftwo}%
\providecommand \translation [1]{[#1]}%
\providecommand \BibitemOpen [0]{}%
\providecommand \bibitemStop [0]{}%
\providecommand \bibitemNoStop [0]{.\EOS\space}%
\providecommand \EOS [0]{\spacefactor3000\relax}%
\providecommand \BibitemShut  [1]{\csname bibitem#1\endcsname}%
\let\auto@bib@innerbib\@empty
\bibitem [{\citenamefont {Reynolds}(1987)}]{reynolds1987flocks}%
  \BibitemOpen
  \bibfield  {author} {\bibinfo {author} {\bibfnamefont {C.~W.}\ \bibnamefont
  {Reynolds}},\ }\href {\doibase 10.1145/37401.37406} {\bibfield  {journal}
  {\bibinfo  {journal} {Comp. Graph.}\ }\textbf {\bibinfo {volume} {21}},\
  \bibinfo {pages} {25} (\bibinfo {year} {1987})}\BibitemShut {NoStop}%
\bibitem [{\citenamefont {Vicsek}\ and\ \citenamefont
  {Zafeiris}(2012)}]{vicsek2012collective}%
  \BibitemOpen
  \bibfield  {author} {\bibinfo {author} {\bibfnamefont {T.}~\bibnamefont
  {Vicsek}}\ and\ \bibinfo {author} {\bibfnamefont {A.}~\bibnamefont
  {Zafeiris}},\ }\href {\doibase 10.1016/j.physrep.2012.03.004} {\bibfield
  {journal} {\bibinfo  {journal} {Phys. Rep.}\ }\textbf {\bibinfo {volume}
  {517}},\ \bibinfo {pages} {71} (\bibinfo {year} {2012})}\BibitemShut
  {NoStop}%
\bibitem [{\citenamefont {Giardina}(2008)}]{giardinarev}%
  \BibitemOpen
  \bibfield  {author} {\bibinfo {author} {\bibfnamefont {I.}~\bibnamefont
  {Giardina}},\ }\href {\doibase 10.2976/1.2961038} {\bibfield  {journal}
  {\bibinfo  {journal} {HFSP J}\ }\textbf {\bibinfo {volume} {2}},\ \bibinfo
  {pages} {205} (\bibinfo {year} {2008})}\BibitemShut {NoStop}%
\bibitem [{\citenamefont {Schaller}\ \emph {et~al.}(2010)\citenamefont
  {Schaller}, \citenamefont {Weber}, \citenamefont {Semmrich}, \citenamefont
  {Frey},\ and\ \citenamefont {Bausch}}]{schaller2010polar}%
  \BibitemOpen
  \bibfield  {author} {\bibinfo {author} {\bibfnamefont {V.}~\bibnamefont
  {Schaller}}, \bibinfo {author} {\bibfnamefont {C.}~\bibnamefont {Weber}},
  \bibinfo {author} {\bibfnamefont {C.}~\bibnamefont {Semmrich}}, \bibinfo
  {author} {\bibfnamefont {E.}~\bibnamefont {Frey}}, \ and\ \bibinfo {author}
  {\bibfnamefont {A.~R.}\ \bibnamefont {Bausch}},\ }\href {\doibase%
  10.1038/nature09312} {\bibfield  {journal} {\bibinfo  {journal} {Nature}\
  }\textbf {\bibinfo {volume} {467}},\ \bibinfo {pages} {73} (\bibinfo {year}
  {2010})}\BibitemShut {NoStop}%
\bibitem [{\citenamefont {Peruani}\ \emph {et~al.}(2012)\citenamefont
  {Peruani}, \citenamefont {Starru\ss{}}, \citenamefont {Jakovljevic},
  \citenamefont {S\o{}gaard-Andersen}, \citenamefont {Deutsch},\ and\
  \citenamefont {B\"ar}}]{peruani2012collective}%
  \BibitemOpen
  \bibfield  {author} {\bibinfo {author} {\bibfnamefont {F.}~\bibnamefont
  {Peruani}}, \bibinfo {author} {\bibfnamefont {J.}~\bibnamefont
  {Starru\ss{}}}, \bibinfo {author} {\bibfnamefont {V.}~\bibnamefont
  {Jakovljevic}}, \bibinfo {author} {\bibfnamefont {L.}~\bibnamefont
  {S\o{}gaard-Andersen}}, \bibinfo {author} {\bibfnamefont {A.}~\bibnamefont
  {Deutsch}}, \ and\ \bibinfo {author} {\bibfnamefont {M.}~\bibnamefont
  {B\"ar}},\ }\href {\doibase 10.1103/PhysRevLett.108.098102} {\bibfield
  {journal} {\bibinfo  {journal} {Phys. Rev. Lett.}\ }\textbf {\bibinfo
  {volume} {108}},\ \bibinfo {pages} {098102} (\bibinfo {year}
  {2012})}\BibitemShut {NoStop}%
\bibitem [{\citenamefont {Guttal}\ \emph {et~al.}(2012)\citenamefont {Guttal},
  \citenamefont {Romanczuk}, \citenamefont {Simpson}, \citenamefont {Sword},\
  and\ \citenamefont {Couzin}}]{guttal2012cannibalism}%
  \BibitemOpen
  \bibfield  {author} {\bibinfo {author} {\bibfnamefont {V.}~\bibnamefont
  {Guttal}}, \bibinfo {author} {\bibfnamefont {P.}~\bibnamefont {Romanczuk}},
  \bibinfo {author} {\bibfnamefont {S.~J.}\ \bibnamefont {Simpson}}, \bibinfo
  {author} {\bibfnamefont {G.~A.}\ \bibnamefont {Sword}}, \ and\ \bibinfo
  {author} {\bibfnamefont {I.~D.}\ \bibnamefont {Couzin}},\ }\href {\doibase%
  10.1111/j.1461-0248.2012.01840.x} {\bibfield  {journal} {\bibinfo  {journal}
  {Ecol. Lett.}\ }\textbf {\bibinfo {volume} {15}},\ \bibinfo {pages} {1158}
  (\bibinfo {year} {2012})}\BibitemShut {NoStop}%
\bibitem [{\citenamefont {Bazazi}\ \emph {et~al.}(2008)\citenamefont {Bazazi},
  \citenamefont {Buhl}, \citenamefont {Hale}, \citenamefont {Anstey},
  \citenamefont {Sword}, \citenamefont {Simpson},\ and\ \citenamefont
  {Couzin}}]{bazazi2008collective}%
  \BibitemOpen
  \bibfield  {author} {\bibinfo {author} {\bibfnamefont {S.}~\bibnamefont
  {Bazazi}}, \bibinfo {author} {\bibfnamefont {J.}~\bibnamefont {Buhl}},
  \bibinfo {author} {\bibfnamefont {J.~J.}\ \bibnamefont {Hale}}, \bibinfo
  {author} {\bibfnamefont {M.~L.}\ \bibnamefont {Anstey}}, \bibinfo {author}
  {\bibfnamefont {G.~A.}\ \bibnamefont {Sword}}, \bibinfo {author}
  {\bibfnamefont {S.~J.}\ \bibnamefont {Simpson}}, \ and\ \bibinfo {author}
  {\bibfnamefont {I.~D.}\ \bibnamefont {Couzin}},\ }\href {\doibase%
  10.1016/j.cub.2008.04.035} {\bibfield  {journal} {\bibinfo  {journal} {Curr.
  Biol.}\ }\textbf {\bibinfo {volume} {18}},\ \bibinfo {pages} {735} (\bibinfo
  {year} {2008})}\BibitemShut {NoStop}%
\bibitem [{\citenamefont {Herbert-Read}\ \emph {et~al.}(2011)\citenamefont
  {Herbert-Read}, \citenamefont {Perna}, \citenamefont {Mann}, \citenamefont
  {Schaerf}, \citenamefont {Sumpter},\ and\ \citenamefont
  {Ward}}]{herbert2011inferring}%
  \BibitemOpen
  \bibfield  {author} {\bibinfo {author} {\bibfnamefont {J.~E.}\ \bibnamefont
  {Herbert-Read}}, \bibinfo {author} {\bibfnamefont {A.}~\bibnamefont {Perna}},
  \bibinfo {author} {\bibfnamefont {R.~P.}\ \bibnamefont {Mann}}, \bibinfo
  {author} {\bibfnamefont {T.~M.}\ \bibnamefont {Schaerf}}, \bibinfo {author}
  {\bibfnamefont {D.~J.}\ \bibnamefont {Sumpter}}, \ and\ \bibinfo {author}
  {\bibfnamefont {A.~J.}\ \bibnamefont {Ward}},\ }\href {\doibase%
  10.1073/pnas.1109355108} {\bibfield  {journal} {\bibinfo  {journal} {Proc.
  Natl. Acad. Sci. U.S.A.}\ }\textbf {\bibinfo {volume} {108}},\ \bibinfo
  {pages} {18726} (\bibinfo {year} {2011})}\BibitemShut {NoStop}%
\bibitem [{\citenamefont {Misund}\ \emph {et~al.}(1995)\citenamefont {Misund},
  \citenamefont {Aglen},\ and\ \citenamefont
  {Fr{\o}n{\ae}s}}]{misund1995mapping}%
  \BibitemOpen
  \bibfield  {author} {\bibinfo {author} {\bibfnamefont {O.~A.}\ \bibnamefont
  {Misund}}, \bibinfo {author} {\bibfnamefont {A.}~\bibnamefont {Aglen}}, \
  and\ \bibinfo {author} {\bibfnamefont {E.}~\bibnamefont {Fr{\o}n{\ae}s}},\
  }\href {\doibase 10.1016/1054-3139(95)80011-5} {\bibfield  {journal}
  {\bibinfo  {journal} {ICES J. Mar. Sci.}\ }\textbf {\bibinfo {volume} {52}},\
  \bibinfo {pages} {11} (\bibinfo {year} {1995})}\BibitemShut {NoStop}%
\bibitem [{\citenamefont {Ballerini}\ \emph {et~al.}(2008)\citenamefont
  {Ballerini}, \citenamefont {N.~Cabibbo}, \citenamefont {Cavagna},
  \citenamefont {Cisbani}, \citenamefont {Giardina},\ and\ \citenamefont
  {Zdravkovic}}]{ballerinitopol}%
  \BibitemOpen
  \bibfield  {author} {\bibinfo {author} {\bibfnamefont {M.}~\bibnamefont
  {Ballerini}}, \bibinfo {author} {\bibfnamefont {R.~C.}\ \bibnamefont
  {N.~Cabibbo}}, \bibinfo {author} {\bibfnamefont {A.}~\bibnamefont {Cavagna}},
  \bibinfo {author} {\bibfnamefont {E.}~\bibnamefont {Cisbani}}, \bibinfo
  {author} {\bibfnamefont {I.}~\bibnamefont {Giardina}}, \ and\ \bibinfo
  {author} {\bibfnamefont {V.}~\bibnamefont {Zdravkovic}},\ }\href {\doibase%
  10.1073/pnas.0711437105} {\bibfield  {journal} {\bibinfo  {journal} {Proc.
  Natl. Acad. Sci. U.S.A.}\ }\textbf {\bibinfo {volume} {105}},\ \bibinfo
  {pages} {1232} (\bibinfo {year} {2008})}\BibitemShut {NoStop}%
\bibitem [{\citenamefont {Pearce}\ \emph {et~al.}(2014)\citenamefont {Pearce},
  \citenamefont {Miller}, \citenamefont {Rowlands},\ and\ \citenamefont
  {Turner}}]{pearce2014role}%
  \BibitemOpen
  \bibfield  {author} {\bibinfo {author} {\bibfnamefont {D.~J.}\ \bibnamefont
  {Pearce}}, \bibinfo {author} {\bibfnamefont {A.~M.}\ \bibnamefont {Miller}},
  \bibinfo {author} {\bibfnamefont {G.}~\bibnamefont {Rowlands}}, \ and\
  \bibinfo {author} {\bibfnamefont {M.~S.}\ \bibnamefont {Turner}},\ }\href
  {\doibase 10.1073/pnas.1402202111} {\bibfield  {journal} {\bibinfo  {journal}
  {Proc. Natl. Acad. Sci. U.S.A.}\ }\textbf {\bibinfo {volume} {111}},\
  \bibinfo {pages} {10422} (\bibinfo {year} {2014})}\BibitemShut {NoStop}%
\bibitem [{\citenamefont {Cavagna}\ \emph {et~al.}(2013)\citenamefont
  {Cavagna}, \citenamefont {Giardina},\ and\ \citenamefont
  {Ginelli}}]{Cavagnaboundaryinfo}%
  \BibitemOpen
  \bibfield  {author} {\bibinfo {author} {\bibfnamefont {A.}~\bibnamefont
  {Cavagna}}, \bibinfo {author} {\bibfnamefont {I.}~\bibnamefont {Giardina}}, \
  and\ \bibinfo {author} {\bibfnamefont {F.}~\bibnamefont {Ginelli}},\ }\href
  {\doibase 10.1103/PhysRevLett.110.168107} {\bibfield  {journal} {\bibinfo
  {journal} {Phys. Rev. Lett.}\ }\textbf {\bibinfo {volume} {110}},\ \bibinfo
  {pages} {168107} (\bibinfo {year} {2013})}\BibitemShut {NoStop}%
\bibitem [{\citenamefont {Bialek}\ \emph {et~al.}(2012)\citenamefont {Bialek},
  \citenamefont {Cavagna}, \citenamefont {Giardina}, \citenamefont {Mora},
  \citenamefont {Silvestri}, \citenamefont {Viale},\ and\ \citenamefont
  {Walczak}}]{bialek2012statistical}%
  \BibitemOpen
  \bibfield  {author} {\bibinfo {author} {\bibfnamefont {W.}~\bibnamefont
  {Bialek}}, \bibinfo {author} {\bibfnamefont {A.}~\bibnamefont {Cavagna}},
  \bibinfo {author} {\bibfnamefont {I.}~\bibnamefont {Giardina}}, \bibinfo
  {author} {\bibfnamefont {T.}~\bibnamefont {Mora}}, \bibinfo {author}
  {\bibfnamefont {E.}~\bibnamefont {Silvestri}}, \bibinfo {author}
  {\bibfnamefont {M.}~\bibnamefont {Viale}}, \ and\ \bibinfo {author}
  {\bibfnamefont {A.~M.}\ \bibnamefont {Walczak}},\ }\href {\doibase%
  10.1073/pnas.1118633109} {\bibfield  {journal} {\bibinfo  {journal} {Proc.
  Natl. Acad. Sci. U.S.A.}\ }\textbf {\bibinfo {volume} {109}},\ \bibinfo
  {pages} {4786} (\bibinfo {year} {2012})}\BibitemShut {NoStop}%
\bibitem [{\citenamefont {Pearce}\ and\ \citenamefont
  {Turner}(2014)}]{pearce2014density}%
  \BibitemOpen
  \bibfield  {author} {\bibinfo {author} {\bibfnamefont {D.~J.}\ \bibnamefont
  {Pearce}}\ and\ \bibinfo {author} {\bibfnamefont {M.~S.}\ \bibnamefont
  {Turner}},\ }\href {\doibase 10.1088/1367-2630/16/8/082002} {\bibfield
  {journal} {\bibinfo  {journal} {New J. Phys.}\ }\textbf {\bibinfo {volume}
  {16}},\ \bibinfo {pages} {082002} (\bibinfo {year} {2014})}\BibitemShut
  {NoStop}%
\bibitem [{\citenamefont {Giomi}\ \emph {et~al.}(2013)\citenamefont {Giomi},
  \citenamefont {Hawley-Weld},\ and\ \citenamefont
  {Mahadevan}}]{giomi2013swarming}%
  \BibitemOpen
  \bibfield  {author} {\bibinfo {author} {\bibfnamefont {L.}~\bibnamefont
  {Giomi}}, \bibinfo {author} {\bibfnamefont {N.}~\bibnamefont {Hawley-Weld}},
  \ and\ \bibinfo {author} {\bibfnamefont {L.}~\bibnamefont {Mahadevan}},\
  }\href {\doibase 10.1098/rspa.2012.0637} {\bibfield  {journal} {\bibinfo
  {journal} {Proc. R. Soc. A}\ }\textbf {\bibinfo {volume} {469}} (\bibinfo
  {year} {2013})}\BibitemShut {NoStop}%
\bibitem [{\citenamefont {Rubenstein}\ \emph {et~al.}(2014)\citenamefont
  {Rubenstein}, \citenamefont {Cornejo},\ and\ \citenamefont
  {Nagpal}}]{rubenstein2014programmable}%
  \BibitemOpen
  \bibfield  {author} {\bibinfo {author} {\bibfnamefont {M.}~\bibnamefont
  {Rubenstein}}, \bibinfo {author} {\bibfnamefont {A.}~\bibnamefont {Cornejo}},
  \ and\ \bibinfo {author} {\bibfnamefont {R.}~\bibnamefont {Nagpal}},\ }\href
  {\doibase 10.1126/science.1254295} {\bibfield  {journal} {\bibinfo  {journal}
  {Science}\ }\textbf {\bibinfo {volume} {345}},\ \bibinfo {pages} {795}
  (\bibinfo {year} {2014})}\BibitemShut {NoStop}%
\bibitem [{\citenamefont {Couzin}\ \emph {et~al.}(2011)\citenamefont {Couzin},
  \citenamefont {Ioannou}, \citenamefont {Demirel}, \citenamefont {Gross},
  \citenamefont {Torney}, \citenamefont {Hartnett}, \citenamefont {Conradt},
  \citenamefont {Levin},\ and\ \citenamefont {Leonard}}]{couzin2011uninformed}%
  \BibitemOpen
  \bibfield  {author} {\bibinfo {author} {\bibfnamefont {I.~D.}\ \bibnamefont
  {Couzin}}, \bibinfo {author} {\bibfnamefont {C.~C.}\ \bibnamefont {Ioannou}},
  \bibinfo {author} {\bibfnamefont {G.}~\bibnamefont {Demirel}}, \bibinfo
  {author} {\bibfnamefont {T.}~\bibnamefont {Gross}}, \bibinfo {author}
  {\bibfnamefont {C.~J.}\ \bibnamefont {Torney}}, \bibinfo {author}
  {\bibfnamefont {A.}~\bibnamefont {Hartnett}}, \bibinfo {author}
  {\bibfnamefont {L.}~\bibnamefont {Conradt}}, \bibinfo {author} {\bibfnamefont
  {S.~A.}\ \bibnamefont {Levin}}, \ and\ \bibinfo {author} {\bibfnamefont
  {N.~E.}\ \bibnamefont {Leonard}},\ }\href {\doibase 10.1126/science.1210280}
  {\bibfield  {journal} {\bibinfo  {journal} {Science}\ }\textbf {\bibinfo
  {volume} {334}},\ \bibinfo {pages} {1578} (\bibinfo {year}
  {2011})}\BibitemShut {NoStop}%
\bibitem [{\citenamefont {Dell'Ariccia}\ \emph {et~al.}(2008)\citenamefont
  {Dell'Ariccia}, \citenamefont {Dell'Omo}, \citenamefont {Wolfer},\ and\
  \citenamefont {Lipp}}]{dell2008flock}%
  \BibitemOpen
  \bibfield  {author} {\bibinfo {author} {\bibfnamefont {G.}~\bibnamefont
  {Dell'Ariccia}}, \bibinfo {author} {\bibfnamefont {G.}~\bibnamefont
  {Dell'Omo}}, \bibinfo {author} {\bibfnamefont {D.~P.}\ \bibnamefont
  {Wolfer}}, \ and\ \bibinfo {author} {\bibfnamefont {H.-P.}\ \bibnamefont
  {Lipp}},\ }\href {\doibase doi:10.1016/j.anbehav.2008.05.022} {\bibfield
  {journal} {\bibinfo  {journal} {Animal Behaviour}\ }\textbf {\bibinfo
  {volume} {76}},\ \bibinfo {pages} {1165} (\bibinfo {year}
  {2008})}\BibitemShut {NoStop}%
\bibitem [{\citenamefont {Ward}\ \emph {et~al.}(2011)\citenamefont {Ward},
  \citenamefont {Herbert-Read}, \citenamefont {Sumpter},\ and\ \citenamefont
  {Krause}}]{ward2011fast}%
  \BibitemOpen
  \bibfield  {author} {\bibinfo {author} {\bibfnamefont {A.~J.}\ \bibnamefont
  {Ward}}, \bibinfo {author} {\bibfnamefont {J.~E.}\ \bibnamefont
  {Herbert-Read}}, \bibinfo {author} {\bibfnamefont {D.~J.}\ \bibnamefont
  {Sumpter}}, \ and\ \bibinfo {author} {\bibfnamefont {J.}~\bibnamefont
  {Krause}},\ }\href {\doibase doi: 10.1073/pnas.1007102108} {\bibfield
  {journal} {\bibinfo  {journal} {Proc. Natl. Acad. Sci. U.S.A.}\ }\textbf
  {\bibinfo {volume} {108}},\ \bibinfo {pages} {2312} (\bibinfo {year}
  {2011})}\BibitemShut {NoStop}%
\bibitem [{\citenamefont {Reebs}(2000)}]{reebs2000can}%
  \BibitemOpen
  \bibfield  {author} {\bibinfo {author} {\bibfnamefont {S.~G.}\ \bibnamefont
  {Reebs}},\ }\href {\doibase 10.1006/anbe.1999.1314} {\bibfield  {journal}
  {\bibinfo  {journal} {Anim. Behav.}\ }\textbf {\bibinfo {volume} {59}},\
  \bibinfo {pages} {403} (\bibinfo {year} {2000})}\BibitemShut {NoStop}%
\bibitem [{\citenamefont {Krause}\ \emph {et~al.}(2000)\citenamefont {Krause},
  \citenamefont {Hoare}, \citenamefont {Krause}, \citenamefont {Hemelrijk},\
  and\ \citenamefont {Rubenstein}}]{krause2000leadership}%
  \BibitemOpen
  \bibfield  {author} {\bibinfo {author} {\bibfnamefont {J.}~\bibnamefont
  {Krause}}, \bibinfo {author} {\bibfnamefont {D.}~\bibnamefont {Hoare}},
  \bibinfo {author} {\bibfnamefont {S.}~\bibnamefont {Krause}}, \bibinfo
  {author} {\bibfnamefont {C.}~\bibnamefont {Hemelrijk}}, \ and\ \bibinfo
  {author} {\bibfnamefont {D.}~\bibnamefont {Rubenstein}},\ }\href {\doibase%
  10.1111/j.1467-2979.2000.tb00001.x} {\bibfield  {journal} {\bibinfo
  {journal} {Fish Fish.}\ }\textbf {\bibinfo {volume} {1}},\ \bibinfo {pages}
  {82} (\bibinfo {year} {2000})}\BibitemShut {NoStop}%
\bibitem [{\citenamefont {Leblond}\ and\ \citenamefont
  {Reebs}(2006)}]{leblond2006individual}%
  \BibitemOpen
  \bibfield  {author} {\bibinfo {author} {\bibfnamefont {C.}~\bibnamefont
  {Leblond}}\ and\ \bibinfo {author} {\bibfnamefont {S.~G.}\ \bibnamefont
  {Reebs}},\ }\href {\doibase 10.1163/156853906778691603} {\bibfield  {journal}
  {\bibinfo  {journal} {Behaviour}\ }\textbf {\bibinfo {volume} {143}},\
  \bibinfo {pages} {1263} (\bibinfo {year} {2006})}\BibitemShut {NoStop}%
\bibitem [{\citenamefont {Faria}\ \emph {et~al.}(2010)\citenamefont {Faria},
  \citenamefont {Dyer}, \citenamefont {Cl\'ement}, \citenamefont {Couzin},
  \citenamefont {Holt}, \citenamefont {Ward}, \citenamefont {Waters},\ and\
  \citenamefont {Krause}}]{faria2010robofish}%
  \BibitemOpen
  \bibfield  {author} {\bibinfo {author} {\bibfnamefont {J.~J.}\ \bibnamefont
  {Faria}}, \bibinfo {author} {\bibfnamefont {J.~R.}\ \bibnamefont {Dyer}},
  \bibinfo {author} {\bibfnamefont {R.~O.}\ \bibnamefont {Cl\'ement}}, \bibinfo
  {author} {\bibfnamefont {I.~D.}\ \bibnamefont {Couzin}}, \bibinfo {author}
  {\bibfnamefont {N.}~\bibnamefont {Holt}}, \bibinfo {author} {\bibfnamefont
  {A.~J.}\ \bibnamefont {Ward}}, \bibinfo {author} {\bibfnamefont
  {D.}~\bibnamefont {Waters}}, \ and\ \bibinfo {author} {\bibfnamefont
  {J.}~\bibnamefont {Krause}},\ }\href {\doibase 10.1007/s00265-010-0988-y}
  {\bibfield  {journal} {\bibinfo  {journal} {Behav. Ecol. Sociobiol.}\
  }\textbf {\bibinfo {volume} {64}},\ \bibinfo {pages} {1211} (\bibinfo {year}
  {2010})}\BibitemShut {NoStop}%
\bibitem [{\citenamefont {Ward}\ \emph {et~al.}(2008)\citenamefont {Ward},
  \citenamefont {Sumpter}, \citenamefont {Couzin}, \citenamefont {Hart},\ and\
  \citenamefont {Krause}}]{ward2008quorum}%
  \BibitemOpen
  \bibfield  {author} {\bibinfo {author} {\bibfnamefont {A.~J.}\ \bibnamefont
  {Ward}}, \bibinfo {author} {\bibfnamefont {D.~J.}\ \bibnamefont {Sumpter}},
  \bibinfo {author} {\bibfnamefont {I.~D.}\ \bibnamefont {Couzin}}, \bibinfo
  {author} {\bibfnamefont {P.~J.}\ \bibnamefont {Hart}}, \ and\ \bibinfo
  {author} {\bibfnamefont {J.}~\bibnamefont {Krause}},\ }\href {\doibase%
  10.1073/pnas.0710344105} {\bibfield  {journal} {\bibinfo  {journal} {Proc.
  Natl. Acad. Sci. U.S.A.}\ }\textbf {\bibinfo {volume} {105}},\ \bibinfo
  {pages} {6948} (\bibinfo {year} {2008})}\BibitemShut {NoStop}%
\bibitem [{\citenamefont {Couzin}\ \emph {et~al.}(2005)\citenamefont {Couzin},
  \citenamefont {Krause}, \citenamefont {Franks},\ and\ \citenamefont
  {Levin}}]{couzin2005effective}%
  \BibitemOpen
  \bibfield  {author} {\bibinfo {author} {\bibfnamefont {I.~D.}\ \bibnamefont
  {Couzin}}, \bibinfo {author} {\bibfnamefont {J.}~\bibnamefont {Krause}},
  \bibinfo {author} {\bibfnamefont {N.~R.}\ \bibnamefont {Franks}}, \ and\
  \bibinfo {author} {\bibfnamefont {S.~A.}\ \bibnamefont {Levin}},\ }\href
  {\doibase 10.1038/nature03236} {\bibfield  {journal} {\bibinfo  {journal}
  {Nature}\ }\textbf {\bibinfo {volume} {433}},\ \bibinfo {pages} {513}
  (\bibinfo {year} {2005})}\BibitemShut {NoStop}%
\bibitem [{\citenamefont {Sumpter}\ \emph {et~al.}(2008)\citenamefont
  {Sumpter}, \citenamefont {Krause}, \citenamefont {James}, \citenamefont
  {Couzin},\ and\ \citenamefont {Ward}}]{sumpter2008consensus}%
  \BibitemOpen
  \bibfield  {author} {\bibinfo {author} {\bibfnamefont {D.~J.}\ \bibnamefont
  {Sumpter}}, \bibinfo {author} {\bibfnamefont {J.}~\bibnamefont {Krause}},
  \bibinfo {author} {\bibfnamefont {R.}~\bibnamefont {James}}, \bibinfo
  {author} {\bibfnamefont {I.~D.}\ \bibnamefont {Couzin}}, \ and\ \bibinfo
  {author} {\bibfnamefont {A.~J.}\ \bibnamefont {Ward}},\ }\href {\doibase%
  10.1016/j.cub.2008.09.064} {\bibfield  {journal} {\bibinfo  {journal} {Curr.
  Biol.}\ }\textbf {\bibinfo {volume} {18}},\ \bibinfo {pages} {1773} (\bibinfo
  {year} {2008})}\BibitemShut {NoStop}%
\bibitem [{\citenamefont {Kao}\ \emph {et~al.}(2014)\citenamefont {Kao},
  \citenamefont {Miller}, \citenamefont {Torney}, \citenamefont {Hartnett},\
  and\ \citenamefont {Couzin}}]{kao2014collective}%
  \BibitemOpen
  \bibfield  {author} {\bibinfo {author} {\bibfnamefont {A.~B.}\ \bibnamefont
  {Kao}}, \bibinfo {author} {\bibfnamefont {N.}~\bibnamefont {Miller}},
  \bibinfo {author} {\bibfnamefont {C.}~\bibnamefont {Torney}}, \bibinfo
  {author} {\bibfnamefont {A.}~\bibnamefont {Hartnett}}, \ and\ \bibinfo
  {author} {\bibfnamefont {I.~D.}\ \bibnamefont {Couzin}},\ }\href {\doibase%
  10.1371/journal.pcbi.1003762} {\bibfield  {journal} {\bibinfo  {journal}
  {PLoS Comput Biol}\ }\textbf {\bibinfo {volume} {10}},\ \bibinfo {pages}
  {e1003762} (\bibinfo {year} {2014})}\BibitemShut {NoStop}%
\bibitem [{\citenamefont {Guttal}\ and\ \citenamefont
  {Couzin}(2011)}]{guttal2011leadership}%
  \BibitemOpen
  \bibfield  {author} {\bibinfo {author} {\bibfnamefont {V.}~\bibnamefont
  {Guttal}}\ and\ \bibinfo {author} {\bibfnamefont {I.~D.}\ \bibnamefont
  {Couzin}},\ }\href {\doibase 10.4161/cib.4.3.14887} {\bibfield  {journal}
  {\bibinfo  {journal} {Commun. Integr. Biol.}\ }\textbf {\bibinfo {volume}
  {4}},\ \bibinfo {pages} {294} (\bibinfo {year} {2011})}\BibitemShut {NoStop}%
\bibitem [{\citenamefont {Vicsek}\ \emph {et~al.}(1995)\citenamefont {Vicsek},
  \citenamefont {Czir{\'o}k}, \citenamefont {Ben-Jacob}, \citenamefont
  {Cohen},\ and\ \citenamefont {Shochet}}]{vicsek}%
  \BibitemOpen
  \bibfield  {author} {\bibinfo {author} {\bibfnamefont {T.}~\bibnamefont
  {Vicsek}}, \bibinfo {author} {\bibfnamefont {A.}~\bibnamefont {Czir{\'o}k}},
  \bibinfo {author} {\bibfnamefont {E.}~\bibnamefont {Ben-Jacob}}, \bibinfo
  {author} {\bibfnamefont {I.}~\bibnamefont {Cohen}}, \ and\ \bibinfo {author}
  {\bibfnamefont {O.}~\bibnamefont {Shochet}},\ }\href {\doibase%
  10.1103/PhysRevLett.75.1226} {\bibfield  {journal} {\bibinfo  {journal}
  {Phys. Rev. Lett.}\ }\textbf {\bibinfo {volume} {75}},\ \bibinfo {pages}
  {1226} (\bibinfo {year} {1995})}\BibitemShut {NoStop}%
\bibitem [{\citenamefont {Czir\'ok}\ \emph {et~al.}(1997)\citenamefont
  {Czir\'ok}, \citenamefont {Stanley},\ and\ \citenamefont
  {Vicsek}}]{czirok1997spontaneously}%
  \BibitemOpen
  \bibfield  {author} {\bibinfo {author} {\bibfnamefont {A.}~\bibnamefont
  {Czir\'ok}}, \bibinfo {author} {\bibfnamefont {H.~E.}\ \bibnamefont
  {Stanley}}, \ and\ \bibinfo {author} {\bibfnamefont {T.}~\bibnamefont
  {Vicsek}},\ }\href {\doibase 10.1088/0305-4470/30/5/009} {\bibfield
  {journal} {\bibinfo  {journal} {J. Phys. A-Math. Gen.}\ }\textbf {\bibinfo
  {volume} {30}},\ \bibinfo {pages} {1375} (\bibinfo {year}
  {1997})}\BibitemShut {NoStop}%
\bibitem [{\citenamefont {Chat\'e}\ \emph {et~al.}(2008)\citenamefont
  {Chat\'e}, \citenamefont {Ginelli}, \citenamefont {Gr\'egoire},\ and\
  \citenamefont {Raynaud}}]{chate2008collective}%
  \BibitemOpen
  \bibfield  {author} {\bibinfo {author} {\bibfnamefont {H.}~\bibnamefont
  {Chat\'e}}, \bibinfo {author} {\bibfnamefont {F.}~\bibnamefont {Ginelli}},
  \bibinfo {author} {\bibfnamefont {G.}~\bibnamefont {Gr\'egoire}}, \ and\
  \bibinfo {author} {\bibfnamefont {F.}~\bibnamefont {Raynaud}},\ }\href
  {\doibase 10.1103/PhysRevE.77.046113} {\bibfield  {journal} {\bibinfo
  {journal} {Phys. Rev. E}\ }\textbf {\bibinfo {volume} {77}},\ \bibinfo
  {pages} {046113} (\bibinfo {year} {2008})}\BibitemShut {NoStop}%
\bibitem [{\citenamefont {Ginelli}(2015)}]{ginelli2015physics}%
  \BibitemOpen
  \bibfield  {author} {\bibinfo {author} {\bibfnamefont {F.}~\bibnamefont
  {Ginelli}},\ }\href {http://arxiv.org/abs/1511.01451} {\bibfield  {journal}
  {\bibinfo  {journal} {arXiv}\ } (\bibinfo {year} {2015})}\BibitemShut
  {NoStop}%
\bibitem [{SIn()}]{SInote}%
  \BibitemOpen
  \href@noop {} {}\bibinfo {note} {See Supplemental Material at  \url{http://wwwhome.lorentz.leidenuniv.nl/~giomi/sup_mat/20151108/} for movies of the self propelled particle model.}
\BibitemShut {Stop}%
\bibitem [{\citenamefont {Loi}\ \emph {et~al.}(2008)\citenamefont {Loi},
  \citenamefont {Mossa},\ and\ \citenamefont {Cugliandolo}}]{loi2008effective}%
  \BibitemOpen
  \bibfield  {author} {\bibinfo {author} {\bibfnamefont {D.}~\bibnamefont
  {Loi}}, \bibinfo {author} {\bibfnamefont {S.}~\bibnamefont {Mossa}}, \ and\
  \bibinfo {author} {\bibfnamefont {L.~F.}\ \bibnamefont {Cugliandolo}},\
  }\href {\doibase 10.1103/PhysRevE.77.051111} {\bibfield  {journal} {\bibinfo
  {journal} {Phys. Rev. E}\ }\textbf {\bibinfo {volume} {77}},\ \bibinfo
  {pages} {051111} (\bibinfo {year} {2008})}\BibitemShut {NoStop}%
\bibitem [{\citenamefont {Wang}\ and\ \citenamefont
  {Wolynes}(2011)}]{wang2011communication}%
  \BibitemOpen
  \bibfield  {author} {\bibinfo {author} {\bibfnamefont {S.}~\bibnamefont
  {Wang}}\ and\ \bibinfo {author} {\bibfnamefont {P.~G.}\ \bibnamefont
  {Wolynes}},\ }\href {\doibase 10.1063/1.3624753} {\bibfield  {journal}
  {\bibinfo  {journal} {J. Chem. Phys.}\ }\textbf {\bibinfo {volume} {135}},\
  \bibinfo {pages} {051101} (\bibinfo {year} {2011})}\BibitemShut {NoStop}%
\bibitem [{\citenamefont {Loi}\ \emph {et~al.}(2011)\citenamefont {Loi},
  \citenamefont {Mossa},\ and\ \citenamefont
  {Cugliandolo}}]{loi2011non-conservative}%
  \BibitemOpen
  \bibfield  {author} {\bibinfo {author} {\bibfnamefont {D.}~\bibnamefont
  {Loi}}, \bibinfo {author} {\bibfnamefont {S.}~\bibnamefont {Mossa}}, \ and\
  \bibinfo {author} {\bibfnamefont {L.~F.}\ \bibnamefont {Cugliandolo}},\
  }\href {\doibase 10.1039/C1SM05819C} {\bibfield  {journal} {\bibinfo
  {journal} {Soft Matter}\ }\textbf {\bibinfo {volume} {7}},\ \bibinfo {pages}
  {10193} (\bibinfo {year} {2011})}\BibitemShut {NoStop}%
\bibitem [{\citenamefont {Berthier}\ and\ \citenamefont
  {Kurchan}(2013)}]{berthier2013non-equilibrium}%
  \BibitemOpen
  \bibfield  {author} {\bibinfo {author} {\bibfnamefont {L.}~\bibnamefont
  {Berthier}}\ and\ \bibinfo {author} {\bibfnamefont {J.}~\bibnamefont
  {Kurchan}},\ }\href {\doibase 10.1038/nphys2592} {\bibfield  {journal}
  {\bibinfo  {journal} {Nat. Phys.}\ }\textbf {\bibinfo {volume} {9}},\
  \bibinfo {pages} {310} (\bibinfo {year} {2013})}\BibitemShut {NoStop}%
\bibitem [{\citenamefont {Szamel}(2014)}]{szamel2014self-propelled}%
  \BibitemOpen
  \bibfield  {author} {\bibinfo {author} {\bibfnamefont {G.}~\bibnamefont
  {Szamel}},\ }\href {\doibase 10.1103/PhysRevE.90.012111} {\bibfield
  {journal} {\bibinfo  {journal} {Phys. Rev. E}\ }\textbf {\bibinfo {volume}
  {90}},\ \bibinfo {pages} {012111} (\bibinfo {year} {2014})}\BibitemShut
  {NoStop}%
\bibitem [{\citenamefont {Levis}\ and\ \citenamefont
  {Berthier}(2015)}]{levis2015single}%
  \BibitemOpen
  \bibfield  {author} {\bibinfo {author} {\bibfnamefont {D.}~\bibnamefont
  {Levis}}\ and\ \bibinfo {author} {\bibfnamefont {L.}~\bibnamefont
  {Berthier}},\ }\href {\doibase 10.1209/0295-5075/111/60006} {\bibfield
  {journal} {\bibinfo  {journal} {Europhys. Lett.}\ }\textbf {\bibinfo {volume}
  {111}},\ \bibinfo {pages} {60006} (\bibinfo {year} {2015})}\BibitemShut
  {NoStop}%
\bibitem [{\citenamefont {Kyriakopoulos}\ \emph {et~al.}(2016)\citenamefont
  {Kyriakopoulos}, \citenamefont {Ginelli},\ and\ \citenamefont
  {Toner}}]{Kyriakopoulos2016leading}%
  \BibitemOpen
  \bibfield  {author} {\bibinfo {author} {\bibfnamefont {N.}~\bibnamefont
  {Kyriakopoulos}}, \bibinfo {author} {\bibfnamefont {F.}~\bibnamefont
  {Ginelli}}, \ and\ \bibinfo {author} {\bibfnamefont {J.}~\bibnamefont
  {Toner}},\ }\href {http://stacks.iop.org/1367-2630/18/i=7/a=073039}
  {\bibfield  {journal} {\bibinfo  {journal} {New J. Phys.}\ }\textbf
  {\bibinfo {volume} {18}},\ \bibinfo {pages} {073039} (\bibinfo {year}
  {2016})}\BibitemShut {NoStop}%
\bibitem [{\citenamefont {Toner}\ and\ \citenamefont
  {Tu}(1995)}]{Toner1995Long}%
  \BibitemOpen
  \bibfield  {author} {\bibinfo {author} {\bibfnamefont {J.}~\bibnamefont
  {Toner}}\ and\ \bibinfo {author} {\bibfnamefont {Y.}~\bibnamefont {Tu}},\
  }\href {\doibase 10.1103/PhysRevLett.75.4326} {\bibfield  {journal} {\bibinfo
   {journal} {Phys. Rev. Lett.}\ }\textbf {\bibinfo {volume} {75}},\ \bibinfo
  {pages} {4326} (\bibinfo {year} {1995})}\BibitemShut {NoStop}%
\bibitem [{\citenamefont {Alicea}\ \emph {et~al.}(2005)\citenamefont {Alicea},
  \citenamefont {Balents}, \citenamefont {Fisher}, \citenamefont
  {Paramekanti},\ and\ \citenamefont {Radzihovsky}}]{alicea2005transition}%
  \BibitemOpen
  \bibfield  {author} {\bibinfo {author} {\bibfnamefont {J.}~\bibnamefont
  {Alicea}}, \bibinfo {author} {\bibfnamefont {L.}~\bibnamefont {Balents}},
  \bibinfo {author} {\bibfnamefont {M.~P.~A.}\ \bibnamefont {Fisher}}, \bibinfo
  {author} {\bibfnamefont {A.}~\bibnamefont {Paramekanti}}, \ and\ \bibinfo
  {author} {\bibfnamefont {L.}~\bibnamefont {Radzihovsky}},\ }\href {\doibase%
  10.1103/PhysRevB.71.235322} {\bibfield  {journal} {\bibinfo  {journal} {Phys.
  Rev. B}\ }\textbf {\bibinfo {volume} {71}},\ \bibinfo {pages} {235322}
  (\bibinfo {year} {2005})}\BibitemShut {NoStop}%
\bibitem [{\citenamefont {Chen}\ \emph {et~al.}(2015)\citenamefont {Chen},
  \citenamefont {Toner},\ and\ \citenamefont {Lee}}]{checn2015critical}%
  \BibitemOpen
  \bibfield  {author} {\bibinfo {author} {\bibfnamefont {L.}~\bibnamefont
  {Chen}}, \bibinfo {author} {\bibfnamefont {J.}~\bibnamefont {Toner}}, \ and\
  \bibinfo {author} {\bibfnamefont {C.~F.}\ \bibnamefont {Lee}},\ }\href
  {http://stacks.iop.org/1367-2630/17/i=4/a=042002} {\bibfield  {journal}
  {\bibinfo  {journal} {New J. Phys.}\ }\textbf {\bibinfo {volume}
  {17}},\ \bibinfo {pages} {042002} (\bibinfo {year} {2015})}\BibitemShut
  {NoStop}%
\bibitem [{\citenamefont {Chen}\ \emph {et~al.}(2016)\citenamefont {Chen},
  \citenamefont {Lee},\ and\ \citenamefont {Toner}}]{chen2016mapping}%
  \BibitemOpen
  \bibfield  {author} {\bibinfo {author} {\bibfnamefont {L.}~\bibnamefont
  {Chen}}, \bibinfo {author} {\bibfnamefont {C.~F.}\ \bibnamefont {Lee}}, \
  and\ \bibinfo {author} {\bibfnamefont {J.}~\bibnamefont {Toner}},\ }\href
  {\doibase 10.1038/ncomms12215} {\bibfield  {journal} {\bibinfo  {journal}
  {Nat. Commun.}\ }\textbf {\bibinfo {volume} {7}} (\bibinfo {year}
  {2016})}\BibitemShut {NoStop}%
\bibitem [{\citenamefont {Attanasi}\ \emph {et~al.}(2014)\citenamefont
  {Attanasi}, \citenamefont {Cavagna}, \citenamefont {Del~Castello},
  \citenamefont {Giardina}, \citenamefont {Grigera}, \citenamefont {Jeli{\'c}},
  \citenamefont {Melillo}, \citenamefont {Parisi}, \citenamefont {Pohl},
  \citenamefont {Shen} \emph {et~al.}}]{attanasi2014information}%
  \BibitemOpen
  \bibfield  {author} {\bibinfo {author} {\bibfnamefont {A.}~\bibnamefont
  {Attanasi}}, \bibinfo {author} {\bibfnamefont {A.}~\bibnamefont {Cavagna}},
  \bibinfo {author} {\bibfnamefont {L.}~\bibnamefont {Del~Castello}}, \bibinfo
  {author} {\bibfnamefont {I.}~\bibnamefont {Giardina}}, \bibinfo {author}
  {\bibfnamefont {T.~S.}\ \bibnamefont {Grigera}}, \bibinfo {author}
  {\bibfnamefont {A.}~\bibnamefont {Jeli{\'c}}}, \bibinfo {author}
  {\bibfnamefont {S.}~\bibnamefont {Melillo}}, \bibinfo {author} {\bibfnamefont
  {L.}~\bibnamefont {Parisi}}, \bibinfo {author} {\bibfnamefont
  {O.}~\bibnamefont {Pohl}}, \bibinfo {author} {\bibfnamefont {E.}~\bibnamefont
  {Shen}},  \emph {et~al.},\ }\href {\doibase 10.1038/nphys3035} {\bibfield
  {journal} {\bibinfo  {journal} {Nat. Phys.}\ }\textbf {\bibinfo {volume}
  {10}},\ \bibinfo {pages} {691} (\bibinfo {year} {2014})}\BibitemShut
  {NoStop}%
\bibitem [{\citenamefont {Attanasi}\ \emph {et~al.}(2015)\citenamefont
  {Attanasi}, \citenamefont {Cavagna}, \citenamefont {Del~Castello},
  \citenamefont {Giardina}, \citenamefont {Jelic}, \citenamefont {Melillo},
  \citenamefont {Parisi}, \citenamefont {Pohl}, \citenamefont {Shen},\ and\
  \citenamefont {Viale}}]{attanasi2015emergence}%
  \BibitemOpen
  \bibfield  {author} {\bibinfo {author} {\bibfnamefont {A.}~\bibnamefont
  {Attanasi}}, \bibinfo {author} {\bibfnamefont {A.}~\bibnamefont {Cavagna}},
  \bibinfo {author} {\bibfnamefont {L.}~\bibnamefont {Del~Castello}}, \bibinfo
  {author} {\bibfnamefont {I.}~\bibnamefont {Giardina}}, \bibinfo {author}
  {\bibfnamefont {A.}~\bibnamefont {Jelic}}, \bibinfo {author} {\bibfnamefont
  {S.}~\bibnamefont {Melillo}}, \bibinfo {author} {\bibfnamefont
  {L.}~\bibnamefont {Parisi}}, \bibinfo {author} {\bibfnamefont
  {O.}~\bibnamefont {Pohl}}, \bibinfo {author} {\bibfnamefont {E.}~\bibnamefont
  {Shen}}, \ and\ \bibinfo {author} {\bibfnamefont {M.}~\bibnamefont {Viale}},\
  }\href {\doibase 10.1098/rsif.2015.0319} {\bibfield  {journal} {\bibinfo
  {journal} {J. R. Soc. Interface}\ }\textbf {\bibinfo
  {volume} {12}},\ \bibinfo {pages} {20150319} (\bibinfo {year}
  {2015})}\BibitemShut {NoStop}%
\end{thebibliography}
\end{document}